\documentclass[12pt, onecolumn, draftclsnofoot]{IEEEtran}
\usepackage{graphicx}
\usepackage{subfigure}

\usepackage{bm}
\usepackage{amsfonts}
\usepackage{amsthm}
\usepackage{amssymb}

\usepackage{color}
\usepackage{mathrsfs} 
\usepackage{flushend}

\usepackage[utf8]{inputenc} 

\usepackage[outdir=./]{epstopdf}

\usepackage{cite}

\usepackage{booktabs}

\ifCLASSINFOpdf
\else
\fi
\usepackage{amsmath}
\usepackage{amsthm}

\usepackage{algorithm}
\usepackage{algorithmic}
\ifCLASSOPTIONcompsoc
 \usepackage[caption=false,font=normalsize,labelfont=sf,textfont=sf]{subfig}
\else
 \usepackage[caption=false,font=footnotesize]{subfig}
\fi
\usepackage{dblfloatfix}

\usepackage{soul}

\begin{document}
%
\title{Vehicular Connectivity on Complex Trajectories: Roadway-Geometry Aware ISAC Beam-tracking}
%
%
%

\author{  Xiao Meng,
          Fan Liu,~\IEEEmembership{Member,~IEEE},
          Christos Masouros,~\IEEEmembership{Senior Member,~IEEE},
          Weijie Yuan,~\IEEEmembership{Member,~IEEE},
          Qixun Zhang,~\IEEEmembership{Member,~IEEE},\\
          and Zhiyong Feng,~\IEEEmembership{Senior Member,~IEEE}
          \thanks{(\emph{Corresponding author: Fan Liu}.)}
          \thanks{X. Meng is with the School of Information and Electronics, Beijing Institute of Technology, Beijing, 100081, China, and is also with the Department of Electrical and Electronic Engineering, Southern University of Science and Technology, Shenzhen 518055, China (email: mengxiao94@bit.edu.cn).}\thanks{F. Liu and W. Yuan are with the Department of Electrical and Electronic Engineering, Southern University of Science and Technology, Shenzhen 518055, China (email: liuf6@sustech.edu.cn; yuanwj@sustech.edu.cn).}
          
          \thanks{C. Masouros is with the Department of Electronic and Electrical Engineering, University College London, London, WC1E 7JE, UK (e-mail: chris.masouros@ieee.org).}

          \thanks{Q. Zhang and Z. Feng are with Key Laboratory of Universal Wireless Communications Ministry of Education, Beijing University of Posts and Telecommunications, Beijing 100876, China (e-mail: zhangqixun@bupt.edu.cn; fengzy@bupt.edu.cn)}
}

\maketitle

\begin{abstract}
In this paper, we propose sensing-assisted beamforming designs for vehicles on arbitrarily shaped roads by relying on integrated sensing and communication (ISAC) signalling.
Specifically, we aim to address the limitations of conventional ISAC beam-tracking schemes that do not apply to complex road geometries.
To improve the tracking accuracy and communication quality of service (QoS) in vehicle to infrastructure (V2I) networks, it is essential to model the complicated roadway geometry.
To that end, we impose the curvilinear coordinate system (CCS) in an interacting multiple model extended Kalman filter (IMM-EKF) framework.
By doing so, both the position and the motion of the vehicle on a complicated road can be explicitly modeled and precisely tracked attributing to the benefits from the CCS.
Furthermore, an optimization problem is formulated to maximize the array gain through dynamically adjusting the array size and thereby controlling the beamwidth, which takes the performance loss caused by beam misalignment into account.
Numerical simulations demonstrate that the roadway geometry-aware ISAC beamforming approach outperforms the communication-only based and ISAC kinematic-only based technique in the tracking performance.
Moreover, the effectiveness of the dynamic beamwidth design is also verified by our numerical results.
\end{abstract}


%
\IEEEpeerreviewmaketitle

\section{Introduction}
%
%
%
%
\IEEEPARstart{T}{he}
emerging autonomous driving applications will necessitate Ultra-low latency Gbps wireless links to ensure road safety\cite{wymeersch20175g}.
In addition, centimetre-level positioning information is also essential to correctly make decisions for driving actions\cite{kuutti2018survey}.
To achieve such goals, vehicle-to-everything (V2X) communication has become one of the key techniques for autonomous driving, of which two potential technologies are widely discussed, namely, dedicated short-range communications (DSRC) \cite{kenney2011dedicated,6193212} and Celluar V2X (C-V2X)\cite{gyawali2020challenges,gonzalez2018analytical}.
While the two techniques do offer basic V2X functionalities, they fall short of the demanding requirements mentioned above.
Indeed, the data rate of DSRC is restricted to 27Mbps,  and the vehicular links may become unreliable in high-density and high-mobility scenarios due to the carrier-sense multiple access mechanism employed\cite{gyawali2020challenges}.
Moreover, the LTE-based system provides Gbps communication service with the positioning accuracy of tens of meters and at a latency often in excess of 1s\cite{wymeersch20175g}.
As for the NR-V2X, the relative position information can be obtained by side-link communication.
Nevertheless, it remains unclear how to accurately acquire the absolute location in real time\cite{gyawali2020challenges}.
Besides, while the positioning error can be reduced to the centimetre-level with the assistance of global navigation satellite-based systems (GNSS), the refresh rate of the positioning information is rather limited\cite{wymeersch20175g}.
Real time localization will necessitate the aid of the road side wireless network, and low latency Gbps links will require accurate formation and steering of high-gain beams.

Thanks to the recent advances in wireless communications, massive multi-input-multi-output (mMIMO) arrays in conjunction with mmWave technologies offer an opportunity to tackle the aforementioned problems\cite{rappaport2013millimeter,heath2016overview}.
In particular, the large bandwidth available at mmWave frequency simultaneously provides a high data rate for communication and potentially high range resolution for sensing.
Meanwhile, the mMIMO array is able to form ``pencil-like" beams accurately steering to the targets of interest, which generates a considerable array gain to compensate for the path-loss incurred by mmWave channels, while enhancing the angle resolution for the sensing functionality.
Critically, the channel of mmWave mMIMO system exhibits sparsity, i.e., there are much fewer Non-Line-of-Sight (NLoS) components compared with the sub-6 GHz band, which is particularly favourable for vehicle localization\cite{ngo2015massive}.
To fully exploit the above advantages of mmWave and mMIMO technologies, it is natural to equip the V2X network with both communication and sensing capabilities, such that the safety and reliability of automated vehicle operation can be significantly improved, with high-speed links and accurate localization performance.
In consideration of all these perspectives, research efforts are well-underway toward the deployment of integrated sensing and communication (ISAC) in V2X networks\cite{gonzalez2016radar,liu2020radar,du2021integrated,yuan2020bayesian,liu2020tutorial,liu2021integrated}.

More relevant to this work, a dedicated radar sensor mounted on a roadside unit (RSU) may assist the communication beam training in vehicle-to-infrastructure (V2I) downlink scenarios\cite{gonzalez2016radar}.
By doing so, the direction finding ability of the radar improves the precision of the beam alignment with significantly reduced beam training overhead, which is however at a cost of extra hardware platform.
To further exploit the performance gain provided in ISAC V2X network, a vehicular beamforming approach without any additional sensor was proposed in \cite{liu2020radar}, which employed the echo of mmWave payload signal to sense the location of the vehicles and predictively constructed the beamformer in high mobility scenarios.
By applying the match-filter or other estimation methods, the vehicle's state parameters can be estimated in real-time.
Then an EKF approach was applied to estimate the present and predict the future position of the vehicle so that an efficient ISAC-based predictive beamforming design can be attained.
On top of that, a message passing based algorithm from a Bayesian perspective was proposed in \cite{yuan2020bayesian} to achieve the same objective.
Overall, compared with the traditional beam training schemes and emerging beam tracking schemes\cite{zhu2018high,liu2019ekf,lim2019beam,tan2021wideband,zhu2017auxiliary}, the ISAC-based predictive beamforming algorithm exhibits the following superiorities:
\begin{itemize}
  \item \textbf{Low tracking overhead.} Different from conventional tracking techniques in \cite{zhu2018high,liu2019ekf,lim2019beam,tan2021wideband,zhu2017auxiliary}, the dedicated pilot is not required in the transmission frame, which benefits from lower latency and continuous data transmission and removes the need for CSI feedback and the associated quantization and feedback errors.
  \item \textbf{Significant matched-filtering gain.} Through ISAC signalling, the whole transmission frame can be used for simultaneous data transmission and sensing, which leads to a higher gain for matched-filtering processing, compared with the conventional beam training/tracking approaches relying on pilots only.
  \item \textbf{Two dimensional localization.} In contrast to conventional approaches operated in the angular domain only, ISAC-based schemes offer additional range information which significantly improves the accuracy of localization.
\end{itemize}

Although numerous studies on ISAC-enabled V2X networks have already been reported in the literature, there are still quite a lot of critical challenges that prevent the practical implementation, where one important issue is the over-simplified assumptions adopted.
While the beam training schemes do not need the prior information of the channel state, the beam tracking algorithms often assume that the targets move around the RSU at a constant absolute velocity\cite{zhu2018high,liu2019ekf} or the target is located at a given angle by treating the variation caused by the movement as noise\cite{lim2019beam,zhu2017auxiliary}, which is obviously mismatched with the realistic scenarios.  
More importantly, in many existing ISAC-based V2X schemes, the vehicle is assumed to drive on a straight road parallel to the antenna array, which represents a corner case and cannot cover the broader scenarios\cite{liu2020radar,yuan2020bayesian,du2021integrated}.
Accordingly, the kinematic model in the existing treatise and the resultant EKF-based beam tracking algorithm only work under limited roadway geometries.
More complicated trajectories would result in severe performance loss for both vehicle tracking and communication due to beam misalignment.

In light of the above, we propose a novel sensing-assisted predictive beamforming scheme in V2I networks that is able to operate on arbitrarily shaped roads, by employing a curvilinear coordinate system (CCS)\cite{jo2016tracking}. 
To further model the complicated driving behavior of the vehicles, multiple kinematic models are taken into considerations. For clarity, the contribution of this paper can be summarized as follows

\begin{itemize}
  \item \textbf{ISAC-based predictive beamforming for arbitrarily shaped road.} Inspired by \cite{wang2002arc,wang2002robust,jo2016tracking}, we model the complicated roadway geometry as a CCS and reveal its relationship with the Cartesian and the polar coordinate systems used for sensing, which facilitates beam tracking for the vehicle with uniform velocity.
  \item \textbf{Interacting multiple model (IMM) based beamforming and vehicle maneuver recognition.}  Based on the CCS, we further consider a scenario where the vehicle intermittently changes its lane. By applying the IMM filter\cite{1988IMM}, a more reliable tracking process is established and the maneuver of the vehicle can be readily identified. 
  \item \textbf{Dynamic beamwidth (DB) scheme under misalignment probability constraint.} To address the positioning and angle uncertainty of the IMM-EKF-based algorithm, we propose a dynamic beamwidth adjustment algorithm. The proposed algorithm is capable of activating dynamic numbers of antennas to optimize the array gain while guaranteeing that the misalignment probability is lower than a given threshold.
\end{itemize}

The remainder of this article is organized as follows, Section II introduces the system model and the modeling of CCS, Section III describes the proposed EKF approach and the corresponding IMM filtering, Section IV proposes the dynamic beamwidth algorithm, Section V provides the numerical results, and finally Section VI concludes the paper.
\\\indent {\emph{Notations}}: Unless otherwise specified, matrices are denoted by bold uppercase letters (i.e., $\mathbf{F}$), vectors are represented by bold lowercase letters (i.e., $\mathbf{x}$), and scalars are denoted by normal font (i.e., $\rho$). 
$\left(\cdot\right)^H$ stands for Hermitian transpose. 
$\arctan\left(\cdot\right)$ and $\arcsin\left(\cdot\right)$ denote inverse tangent and inverse sine function in radian. 
$\lfloor n \rfloor$ denote the maximum integer no larger than a real number $n$ , and $\mathbb{E}\{\cdot\}$ represent the statistical expectation.
$\{x,y\}$ and $(s,n)$ denotes the location in the Cartesian coordinate system and the CCS, respectively.

\section{System Model}
In this paper, we consider a V2I downlink scenario where an RSU equipped with a massive antenna array and working at mmWave frequency servers a single antenna vehicle.
The vehicle is assumed to be driven on a road with an arbitrary shape on the $x$-$y$ plane, and the RSU communicates with the vehicle via Line-of-Sight (LoS) channel. Moreover, a uniform planner array (UPA) with $M$ columns and $N$ rows is assumed to be deployed at the RSU.
The shape of the road around the RSU is known by importing a map with high precision or acquiring the roadway information from the road designers. 
Given the page limit, we designate the discussion of the NLoS channel and the road in 3-dimensional models as our future work.

\subsection{Road Geometry Model}
Our goal is to track the variation of the position and the motion of the vehicle so that we can use quite a narrow beam to communicate with the vehicle and obtain the high array gain.
In contrast to the straight road model considered in \cite{liu2020radar,yuan2020bayesian,du2021integrated}, the kinematic equation of the vehicle on a curve road is extremely difficult to be described either in the Cartesian coordinate system or in the polar coordinate system.
This motivates us to model the motion of the vehicle on such road in CCS.
\subsubsection{Basic Concept of Curvilinear Coordinate System}
\begin{figure}[htbp]
  \centering
  \begin{minipage}[t]{0.48\textwidth}
    \centering
    \subfigure []{
    \includegraphics [width=0.30\textwidth]{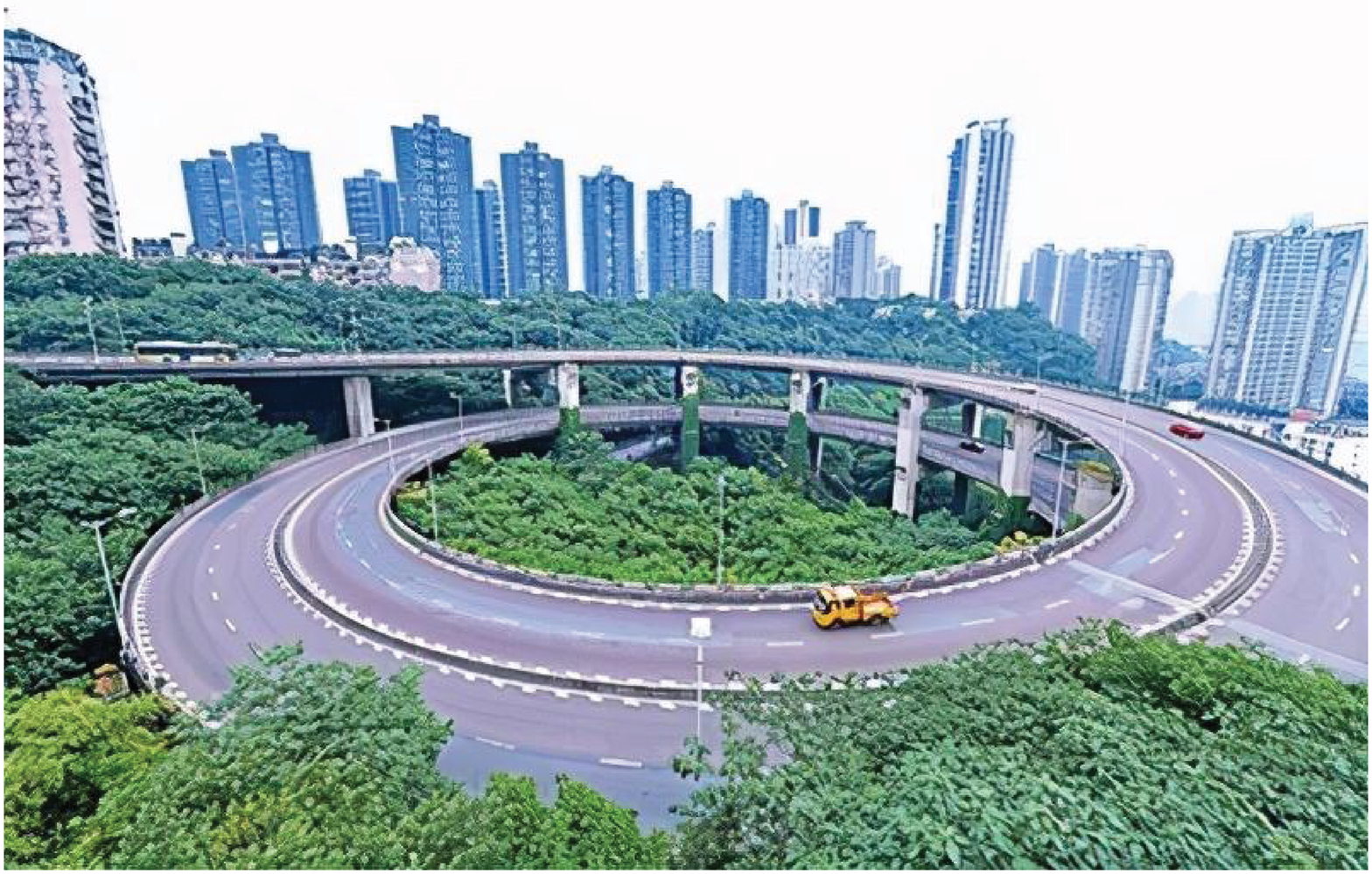}}
    \subfigure []{
    \includegraphics [width=0.30\textwidth]{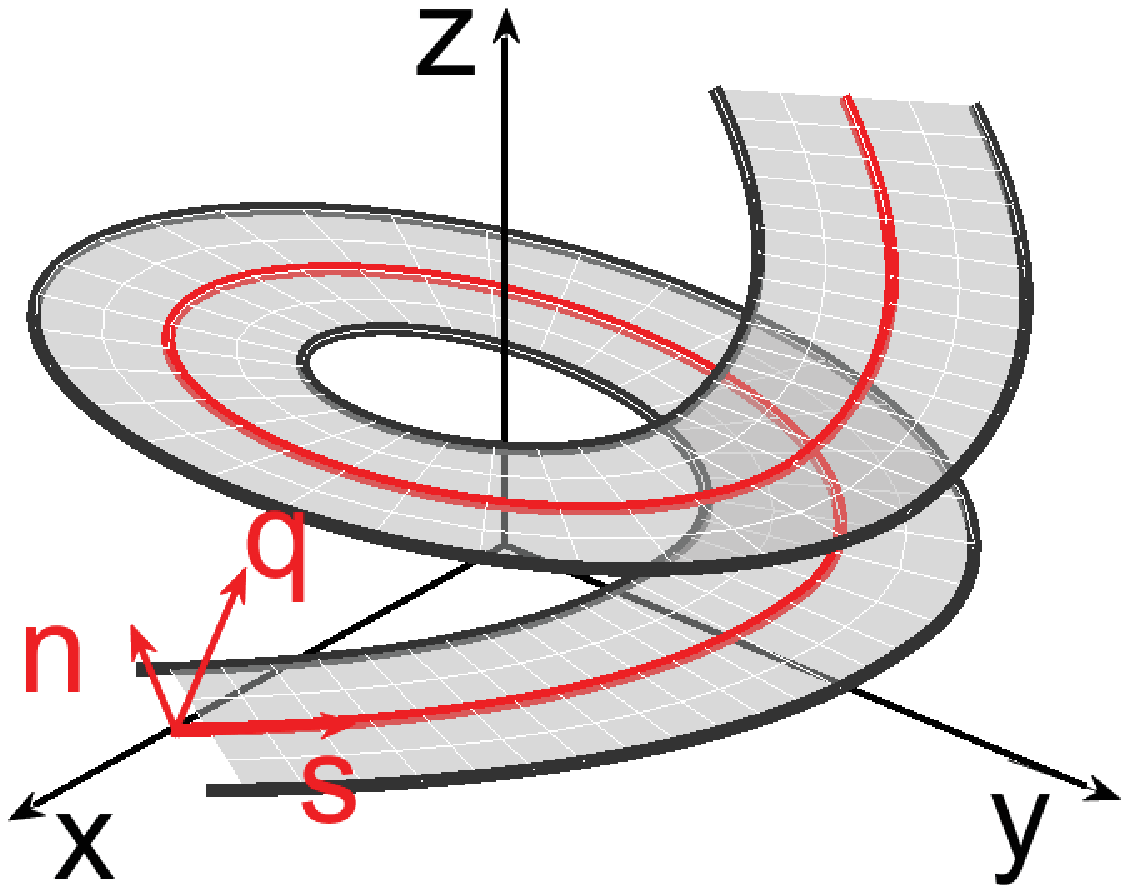}}
    \subfigure []{
    \includegraphics [width=0.30\textwidth]{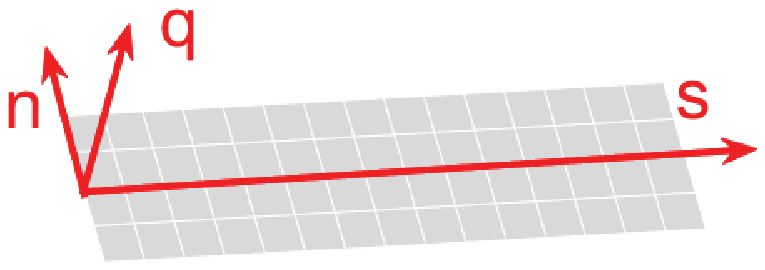}}  
    \caption{The Location and motion can be easily modeled in Curvilinear Coordinate system}\label{CurviDemo}
  \end{minipage}
\quad
  \begin{minipage}[t]{0.48\textwidth}
    \centering
    \includegraphics[width=1\textwidth]{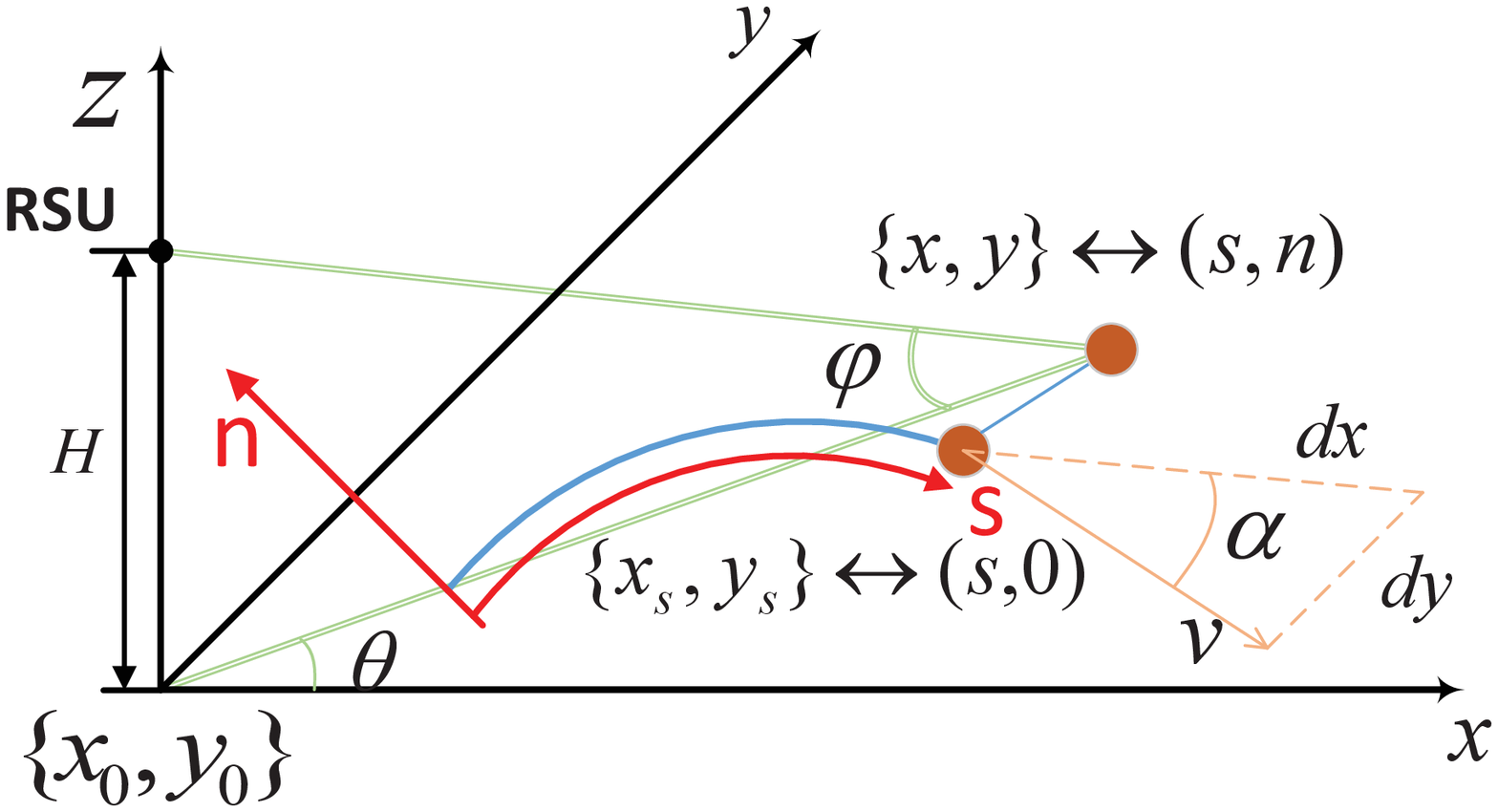}
    \caption{The relationship between the CCS, Cartesian coordinate system and the polar system.}\label{Coordinate}
  \end{minipage}
  \end{figure}

The CCS is characterized in the Euclidean space where a certain number of the coordinate axes are represented based on a curved geometry. As shown in Fig. \ref{CurviDemo}, the complicated road described by $x$, $y$, and $z$ in the Cartesian coordinate system can be represented in CCS by using $s$ and $n$, where the surface of the road is assumed to be flat and thus the height of the road in CCS is always $q=0$.

Based on the CCS, the position and motion have more clear meaning compared with that in the Cartesian coordinate system. 
The $s$ axis represents the dimension that follows the curvature of the arbitrary road, with the value of $s$ denoting the distance travelled along that axis. The $n$ axis is perpendicular to $s$ and the value of $n$ represents the lateral distance from the $s$-axis. The $q$ axis is perpendicular to $s$ and $n$ and the values of $q$ represent the distance on the height dimension in our scenario.
By converting the kinematic description into the roadway-geometry-based CCS, the motion of the vehicle can be easily described in an explicit way.
\subsubsection{Description of the Curve by Fitting Equations}

To set up such a coordinate system under road geometry, we need to first choose a proper method to describe the axis of the new coordinate system.
Without the loss of generality, we use the cubic spline interpolation algorithm in this paper to fit the line that runs across the middle of the road by using a set of parametric functions. Accordingly, the new coordinate system can be described by this set of parameters.
The cubic spline interpolation algorithm can offer not only a high fitting precision but also continuous second order derivation, which is a usefull property in the state description for our EKF in the following.

In order to establish the cubic spline interpolation algorithm, $i+1$ control points should be designated, which include the start point and the endpoint of the road. The other points should also be carefully chosen obeying the following criterion: the larger the curvature is, the more control points are needed.
Besides, since the road designers sometimes use the cubic spline interpolation algorithm to describe the curve, the fitting algorithm can achieve fairly high accuracy if the control points used by the road designer are available.

\subsubsection{Interplay between the Coordinate Systems}
In the considered ISAC system, we estimate the classic sensed parameters, such as Doppler frequency and angle of arrival (AoA), which are naturally described in the polar coordinate system.
Thus, the relationship between the different coordinate systems should be revealed.

As shown in Fig. \ref{Coordinate}, the RSU is located at $\{x_0,y_0\}$ with the height being $H$ to give a bird-eye view. $\theta$ and $\varphi$ represent the azimuth angle and elevation angle, respectively.
Since the conversion between the polar and Cartesian coordinate system is straightforward, we mainly focus on the relationship between the CCS and the Cartesian coordinate system.
By applying the cubic spline interpolation algorithm, the parametric equations of the $i$-th segment of the road between the $i$-th and the $(i+1)$-th control points can be represented as:
\begin{equation}\label{parametric}
  [x_s,y_s,s]^{T} = [\bm{a}_i,\bm{b}_i,\bm{c}_i]^{T} [\rho^3,\rho^2,\rho,1]^{T}
\end{equation}
where $s$ is the longitudinal distance from the start point to the current point in CCS, $\{x_s$,$y_s\}$ is the position of the central point of the road with the distance $s$, i.e., $(s,0)$ in the CCS,  $\bm{a}_{i}$, $\bm{b}_{i}$, $\bm{c}_{i}\in \mathbb{R}^{4\times1}$ are the parameter sets which calculated by the cubic spline interpolation algorithm in the $i$-th segment, and $\rho$ is the variable of the parametric functions.

Following the definition of the longitudinal and lateral distances, the direction of $\vec{n}$ is always vertical to the tagent line of the road at the point $(s,0)$. Accordingly, the point of interest $(s,n)$ can be expressed in the Cartesian coordinate system by the following equations:
\begin{equation}
  [x,y] = [x_s,y_s] + n[-\sin\alpha,\cos\alpha],\label{x_y}
\end{equation}
where $\alpha$ is the angle of the tagent line at $(s,0)$, which can be expressed as
\begin{equation}
  \alpha = \arctan {\frac{dy_s}{dx_s}}\label{def_alpha},
\end{equation}
where $dy_s/dx_s$ can be easily calculated from (\ref{parametric}).
Now, any given point  $(s,n)$ in the curvilinear coordinate system can be converted to the Cartesian coordinate system following the equaitons above and the conversion from the Cartesian coordinate system to the CCS is similar.

\subsection{Signal Model}
\begin{figure}[!htbp]
\centering
\includegraphics[width=0.42\textwidth]{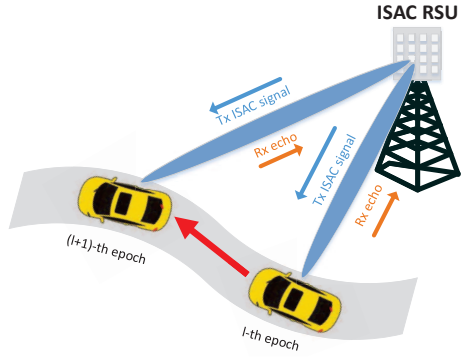}
\caption{ISAC V2I scenario model.}\label{ISAC_Fig}
\end{figure}

As shown in Fig. \ref{ISAC_Fig}, the RSU transmits the ISAC signal to the vehicle and receives the echo signal from the vehicle simultaneously in our considered ISAC V2I downlink scenario.
In each epoch, the RSU tries to construct a beam which steers to the vehicle by applying the predicted angle of the vehicle.
To explore how the transmit and receive signal impact the sensing and communication performance, we investigate the signal model in this subsection.
Since the initial state of the vehicles can be easily obtained by conventional device-based sensing methods, such as beam training and handoff between the different RSUs, we focus on the signal model of the tracking state at the $l$-th epoch.

\subsubsection{Communication  Signal Model}
Let us denote the downlink transmitted ISAC data stream at the $l$-th epoch and time $t$ as $\bar{s}_l(t)$.
The transmit signal can be expressed as 
\begin{equation}
  \tilde{\mathbf{s}}_l(t) = \sqrt{p_l}\mathbf{f}_l \bar{s}_l(t)\in \mathbb{C}^{N_t \times 1},
\end{equation}
where $\mathbf{f}_l$ is the transmit beamforming, with $N_t = M \times N$ being the size of the transmit antenna array and $p_l$ is the total transmit power at the RSU.
Since we consider a LoS channel between the RSU and the vehicle with a single antenna, the $m-$th entry of the array steering vector can be modeled as
\begin{equation}
  a(\theta,\varphi)_{m} =  
  e^{-j\pi (\lfloor\frac{m}{M} \rfloor \sin{\varphi}\cos{\theta}+
    (m-M\lfloor\frac{m}{M} \rfloor) \sin{\varphi}\sin{\theta})}. 
\end{equation}
As we aim to acheive the highest array gain, the transmit beamformer should be designed in the follwoing form, which corresponds to the channel vector,
\begin{equation}
  \mathbf{f}_l = \mathbf{a}(\theta_{l},\varphi_{l}) \approx \mathbf{a}(\hat{\theta}_{l|l-1},\hat{\varphi}_{l|l-1})\label{beampattern},
\end{equation}
where $\hat{\theta}_{l|l-1}$ and  $\hat{\varphi}_{l|l-1}$ are the predicted angles at the $l$-th epoch based on the estimation at the $(l\!-\!1)$-th epoch since the real angle is unavailable to the RSU.

The receive signal at the vehicle can be accordingly expressed as
\begin{equation}
  r_{c,l}(t) = \sqrt{p_l \alpha_l }  \mathbf{a}^H(\theta_{l},\varphi_{l}) \tilde{\mathbf{s}}(t-\tau_{c,l}) e^{j2\pi\varrho_{c,l}t} + z_{c}(t),
\end{equation}
with $\alpha_l$, $\tau_{c,l}$ and $\varrho_{c,l}$ being the path-loss, the time delay and the Doppler frequency at the vehicle at the $l$-th epoch, respectively. \footnote{The predicted Doppler frequency and delay can be estimated by the RSU and fed to the vehicular user (VU) alleviate the computational overhead of the VU.}
Assuming the RSU is equipped with omnidirectional antennas, the path-loss can be modeled as\cite{richards2005fundamentals}
\begin{equation}
  \alpha_{l}(\mbox{dB}) \!=\!  32.4 \!+\! 20\log_{10}f_c(\mbox{MHz}) \!+\! (20 \! \times\! \eta)\log_{10}d_l(\mbox{km}),
\end{equation}
with $\eta$ being the path-loss factor corresponding to the carrier frequency and the electromagnetic propagation environment and we assume $\eta=1$ in this paper.

\subsubsection{Radar Signal Model}
Compared with the communication signal, the Doppler frequency and time delay of the radar signal are doubled due to the round trip of the reflected signal. 
The power of the reflected signal is not only determined by the round-trip path loss but also by the radar cross-section (RCS) of the target.
Thus, the received signal vector can be expressed in the following form
\begin{align}\label{Radar-signal}
  \mathbf{r}_l(t) = \sqrt{p_{l}} \beta_{l} \mathbf{b}(\theta_{l},\varphi_{l}) \mathbf{a}^H(\theta_{l},\varphi_{l}) \tilde{\mathbf{s}}(t-\tau_{l}) e^{j2\pi\mu_{l}t} + \mathbf{z}_r(t),
\end{align}
where $\mathbf{z}_r \in \mathbb{C}^{N_r \times 1}$ denotes the complex additive white Gaussian noise with zero mean and variance of $\sigma ^2$ and $\beta_{l}$, $\mu_{l}$ and $\tau_{l}$ represent the reflection coefficient, the Doppler frequency and the time delay for the RSU.
Besides, the receive channel vector $\mathbf{b}(\theta,\varphi)$ has the same form as the transmit channel vector, i.e., $\mathbf{b}(\theta,\varphi) = \mathbf{a}(\theta,\varphi)$.
Based on the radar equation, the reflection coefficient of this ISAC system with standard omnidirectional antennas, which contains the RCS and path-loss, can be modeled as\cite{richards2005fundamentals}
\begin{equation}
  \beta_l = \frac{\lambda \epsilon_l}{(4\pi)^{3/2} d_l^{(2\times \eta)}},
\end{equation}
where $\epsilon_l$ represents the RCS at the $l$-th epoch, $d_l$ represents the propagation distance and $\lambda$ is the wavelength of the carrier, which can be mathematically expressed as $\lambda = f_c/c$ with $f_c$ being the frequency of carrier and $c$ being the speed of the light.
Again, we assume a free space propagation which leads to $\eta = 1$.
\section{The Proposed Approach}
\subsection{Extended Kalman Filter}\label{EKF}
\subsubsection{State Evolution Model}\label{Evolution-Model}
To accurately track the position and the reflection coefficient of the vehicle, the kinematic states should be properly modeled.
Based on the aforementioned curvilinear coordinate system, the position state and the motion state of the vehicle can be naturally decomposed into two orthogonal directions, namely the longitudinal and lateral directions.

In this paper, we assume that the velocity of the vehicle remains unchanged within a single epoch. So, without considering the system noise, the distance and the velocity in the two directions at the $l-$th epoch can be expressed by
$s_l= s_{l-1} + v_{s,l-1}\Delta T$,$n_l= n_{l-1} + v_{n,l-1}\Delta T$, $v_{s,l}=           v_{s,l-1}$ and $v_{n,l} = v_{n,l-1}$,
where $\Delta T$ is the duration of a single epoch.
By defining the propagation distance of the signal as
\begin{equation}
  d = \sqrt{(x_0 - x)^2 + (y_0 - y)^2 + z_0^2},
\end{equation}
the reflection coefficient in the two adjacent epoch can be expressed as
\begin{equation}
  \beta_l = \frac{\lambda \epsilon_l}{(4\pi)^{3/2} d_l^2},  \quad
  \beta_{l-1} = \frac{\lambda \epsilon_{l-1}}{(4\pi)^{3/2} d_{l-1}^2},
\end{equation}
where $\{x_0,y_0,z_0\}$ and $\{x,y\}$ denote the Cartesian coordinates of the RSU and the vehicle, respectively, and $\varepsilon_{l}$ and $\varepsilon_{l-1}$ denote the complex RCS at the $l$-th epoch and $l\!-\!1$-th epoch, respectively.
The RCS of the vehicle is assumed to be a constant within a short period, i. e., $\varepsilon_{l}\approx \varepsilon_{l-1}$, corresponding to a Swerling I target model \cite{richards2005fundamentals}.
Therefore, the evolution of the reflection coefficient can be formulated as
\begin{equation}
  \beta_{l} = \beta_{l-1} \frac{\varepsilon_l d_{l-1}^2}{\varepsilon_{l-1} d_{l}^2}
            \approx \beta_{l-1} \frac{d_{l-1}^2}{d_{l}^2}.
\end{equation}
The full state evolution model can  be accordingly summarized as
\begin{equation}\label{State-Model}
  \begin{cases} 
  s_l     = s_{l-1} + v_{s,l-1}\Delta T +\omega_s,\\ 
  v_{s,l} =           v_{s,l-1}         +\omega_{v_s},\\
  n_l     = n_{l-1} + v_{n,l-1}\Delta T +\omega_n,\\
  v_{n,l}  =           v_{n,l-1}         +\omega_{v_n},\\
  \beta_{l}= \beta_{l-1} \frac{d_{l-1}^2}{d_{l}^2} + \omega_{\beta}
  \end{cases}
\end{equation}
where $\omega_s$, $\omega_{v_s}$, $\omega_n$, $\omega_{v_n}$ and $\omega_{\beta}$ are the system noise of the state evolution function, which are related to the approximation and systematic error of the description to the system.

Further, we introduce two classic motion models to interpret the behavior of the vehicle.
When the vehicle is moving stably on the road at a constant velocity without any turning, we can describe the motion of the vehicle by using $s$, $n$ and $v_s$.
In this model, the lateral velocity and its corresponding system noise are considered as $v_s=0$, $\omega_{v_s}=0$, which is called lane-keeping (LK).
On the other hand, in the lane-changing (LC) model, we describe the motion of the vehicle by taking the lateral velocity $v_n$ into account when the vehicle is changing the lane. 
By adding such an additional state variable, the maneuver of the vehicle can be described while the additional system noise is also introduced.

\subsubsection{Radar Measurement Model}
Once the RSU receives the echo signal at the $l$-th epoch, we can firstly invoke the classic MUltiple SIgnal Classification (MUSIC) algorithm to estimate the AoA, $\hat{\theta_{l}}$ and  $\hat{\varphi_{l}}$\cite{1989MUSIC}.
Then, by employing the Angle and Phase EStimation (APES) algorithm\cite{Xu2008Target}, the reflection coefficient $\hat{\beta_{l}}$ can also be readily obtained.
After that, by constructing the receive beamformer from the estimated AoA, the weighted received signal can be expressed as
\begin{align}
  \tilde{r}_{l}(t)= \sqrt{p_{l}} \beta_{l} 
  \mathbf{w}^H(\hat{\theta}_{l},\hat{\varphi}_{l}) 
  \mathbf{b}(\theta_{l},\varphi_{l}) 
  \mathbf{a}^H(\theta_{l},\varphi_{l}) 
  \tilde{\mathbf{s}}(t-\tau_{l}) e^{j2\pi\mu_{l}t} + \tilde{z}_r(t),
\end{align}
where $\mathbf{w}^H(\theta_{l},\varphi_{l}) $ is the receive beamforming vector which can be expressed as
\begin{equation}
  \mathbf{w}(\hat{\theta}_{l},\hat{\varphi}_{l}) =\sqrt{\frac{1}{N_r}}\mathbf{b}(\hat{\theta}_{l},\hat{\varphi}_{l}).
\end{equation}
It should be highlighted that by using the massive MIMO array at the RSU, the transmit and receive beams are sufficiently narrow, such that the inter-beam interference can be omitted for multiple vehicles scenario. This follows from the established mMIMO theory and can be mathematically expressed as follow\cite{ngo2015massive}:
\begin{equation}
  |a^H(\theta_0,\varphi_0)a(\theta_1,\varphi_1)| \to 0, \forall \theta_0 \neq \theta_1 \,\text{or}\, \varphi_0 \neq \varphi_1, N_t \to \infty.
\end{equation}
Then, by employing matched-filter with a delayed and Doppler-shifted counterpart of $s_{l}(t)$, one can estimate the delay $\tau_{l}$ and the Doppler frequency $\mu_{l}$, which can be anlaytically given as
\begin{equation}
  \{\hat{\tau}_{l},\hat{\mu}_{l}\}  = \arg \max_{\tau,\mu}\left\lvert \int_{0}^{\Delta T} \tilde{r}_{l}(t) \bar{s}_{l}(t-\tau) e^{-j2\pi\mu t} \,dt \right\rvert ^2.
\end{equation}

For the sake of simplicity, we model the estimation error of the unbiased estimator as additive white Gaussian noise with zero mean\cite{1989MUSIC,richards2005fundamentals,Erik2004Book}, which indicates that the parameter estimates obey the following equations:
\begin{equation}
  \label{MES-Model}
  [\hat{\theta}_{l},\hat{\varphi}_{l},\hat{\mu}_{l},\hat{\tau}_{l},\hat{\beta}_{l}] = 
  [\theta_{l},\varphi_{l},\frac{2\cos(\varphi)v_{R,l}}{\lambda},\frac{2d_{l}}{c},\beta_{l}] + \mathbf{z}
\end{equation}
where $v_{R,l}$ is the radial velocity which can be expressed as $v_{R,l} = v_{s,l}\cos(\theta-\alpha) + v_{n,l} \sin(\theta-\alpha)$, $\mathbf{z}$ = [$z_{\theta}$, $z_{\varphi}$, $z_{\mu}$, $z_{\tau}$, $z_{\beta}$] is measurement noise vector.

\subsubsection{Covariance Matrices Approximation}
To better exploit the EKF algorithm, the covariance matrices of the measurement noise should be properly set.
Before analyzing the covariance matrices, we firstly rewrite the received signal in a more compact form to derive the receive signal to noise ratio (SNR), which has an explicit impact on the measurement noise, and can be given by
\begin{equation}
  \tilde{r}_{l}(t) = \sqrt{p_{l}}\beta_{l} \kappa_{T,l} \kappa_{R,l} \bar{s}_{l}(t-\tau_{l}) + \tilde{z}_r(t), \label{Radar-Receive-k}
\end{equation}
where $\kappa_{T}$ and $\kappa_{R}$ are the transmit and receive beamforming gain which are defined as 
\begin{align}
  \kappa_{R,l} =&   \sqrt{\frac{1}{N_r}} \mathbf{w}^H(\hat{\theta}_{l},\hat{\varphi}_{l}) \mathbf{b}(\theta_{l},\varphi_{l}) \leq \sqrt{N_r},\\
  \kappa_{T,l} =&  \sqrt{\frac{1}{N_t}} \mathbf{a}^H(\theta_{l},\varphi_{l})                 \mathbf{a}(\hat{\theta}_{l|l-1},\hat{\varphi}_{l|l-1})\leq \sqrt{N_t}.\label{TxGain}
\end{align}
Thus, the per antenna receive SNR and the beamformed SNR can be expressed as
\begin{equation}
 \rho_0 = \frac{p_{l} \beta^2_{l} \kappa^2_{T,l} }{\sigma^2},
\quad 
  \rho_1 = \frac{p_{l}\beta^2_{l} \kappa^2_{T,l} \kappa^2_{R,l}}{\sigma^2}.
\end{equation}

Since what we actually estimated by applying MUSIC algorithm are the spatial frequencies $a = \pi\sin\varphi\cos\theta$ and $b = \pi\sin\varphi\sin\theta$, the actual amizuth angle and elevation angle and the corresponding CRLB can be calculated by the following equaitons,
\begin{align}
  &\theta = \arctan{\left(\frac{b}{a}\right)}, \qquad \varphi= \arcsin{\frac{\sqrt{a^2+b^2}}{\pi}},\\
 \label{CRLB_angle}
&\mathbf{C}_{\theta,\varphi}
=
\left[\begin{matrix}
  &\frac{\partial \theta}{\partial a}
  &\frac{\partial \theta}{\partial b}\\
  &\frac{\partial \varphi}{\partial a}
  &\frac{\partial \varphi}{\partial b}
\end{matrix}\right]
\left[\begin{matrix}
  &C_a &0\\
  &0   &C_b
\end{matrix}\right]
\left[\begin{matrix}
  &\frac{\partial \theta}{\partial a}
  &\frac{\partial \theta}{\partial b}\\
  &\frac{\partial \varphi}{\partial a}
  &\frac{\partial \varphi}{\partial b}
\end{matrix}\right]^T,
\end{align}
where the Jacobian matrix is easy to calculate and is thus omitted. The CRLB of $a$ and $b$ can be formulated as
\begin{equation}
C_a = C_b = \frac{6}{N_{Sample}^2 N_r \rho_0},
\end{equation}  
with $N_{sample}$ being the number of signal samples.
As for the reflection coefficient, while the CRLB of APES algorithm has not been explicity given, we apply the empirical equation to estimate CRLB of which the approximated error can be verified to be quite small\cite{Erik2004Book}.
Thus, we approximate the lower bound of the estimation error of $\beta$ as
\begin{equation}\label{CRLB_beta}
  C_{\beta} = \frac{1}{N_{sample}N_r \rho_0}.
\end{equation}
The CRLB of estimation for a classic MF algorithm can be directly expressed as \cite{richards2005fundamentals}
\begin{align}
  C_{\tau} = \frac{3}{2\pi^2B_w\chi},  C_{\mu} = \frac{1}{(2\pi)^2\Delta T^2 f_s \chi}\label{CRLB_mu}    . 
\end{align}
In (\ref{CRLB_mu}), $B_w$ and $f_s$ denote the bandwith of the ISAC signal and the sample rate of the analog-digital converter (ADC), respectively, and $\chi$ is the output power SNR of the MF with the following defination
\begin{equation}
  \chi = \frac{E_s}{N_0} = \frac{P_r \Delta T B_w}{\sigma^2} = \rho_1\Delta T B_w.
\end{equation}

\subsubsection{Standard  EKF Procedure}
In this subsection, we propose a Kalman filtering scheme under the road geometry constraint for beam predicting and tracking. 
Although the nonlinearity of the kinematic model is alleviated by applying the CCS, the nonlinearity of the conversion between the polar coordinate system and the CCS still makes it impossible to directly employ the linear Kalman filter in our system.
Thus, we consider an extended Kalman filtering approach that employs the first-order Taylor expansion as the approximation of the nonlinear model.
By denoting the state variables and the measurement signal vector as  $\mathcal{X} =[s, v_s, n, v_n,\beta]^T$  and $\mathcal{Y}  = [\hat{\theta},\hat{\varphi},\hat{\mu},\hat{\tau},\hat{\beta}]$, respectively, the model derived in (\ref{State-Model}) and (\ref{MES-Model}) can be recast in compact forms as
\begin{align}
  \begin{cases}
    & \text{State Evolution Model: } \mathcal{X}_l = \mathbf{g}(\mathcal{X}_{l-1}) + \bm{\omega}_{l},\\
    & \text{Measurement Model: }     \mathcal{Y}_l = \mathbf{h}(\mathcal{X}_{l})   + \mathbf{z}_{l},
  \end{cases}
\end{align}
where $\mathbf{g}(\cdot)$ is defined in (\ref{State-Model}), with $\bm{\omega}\!=\![\omega_s,\!\omega_{v_s},\!\omega_n,\!\omega_{v_n},\!\omega_{\beta}]^T$ being the system noise vector 
and $\mathbf{h}(\cdot)$ is defined in (\ref{MES-Model}) with $\mathbf{z} = [z_{\theta},z_{\varphi},z_{\mu},z_{\tau},z_{\beta}]^T$ being the measurement noise vector.
As considered above, both $\bm{\omega}$ and $\mathbf{z}$ can be modeled as zero-mean Gaussian distributed noise with the covariance matrices being expressed as 
\begin{equation}
  \mathbf{Q}_s = 
  \begin{bmatrix}
    \sigma_{s}^2        & \sigma_{s,v_s}^2  & 0               & 0                 & 0\\
    \sigma_{s,v_s}^2    & \sigma_{v_s}^2    & 0               & 0                 & 0\\
    0                   & 0                 & \sigma_{n}^2    &\sigma_{n,v_n}^2   & 0\\
    0                   & 0                 & \sigma_{n,v_n}^2& \sigma_{v_n}^2     & 0\\
    0                   & 0                 & 0               &0                  &\sigma_{\beta}^2
  \end{bmatrix},
  \mathbf{Q}_m = \text{diag}(C_{\theta,\varphi},C_{\tau},C_{\mu},C_{\beta}).
\end{equation}
To linearize the models, we give the the Jacobian matrices for both $\mathbf{g}(\mathcal{X})$ and $\mathbf{h}(\mathcal{X})$ as
\begin{equation}
  \frac{\partial{\mathbf{g}}}{\partial{\mathcal{X}}} = 
  \begin{bmatrix}
  1 &0 &\Delta T  &0       &0\\
  0 &1 &0         &\Delta T&0\\
  0 &0 &1         &0       &0\\
  0 &0 &0         &1       &0\\
  \frac{\partial \check{\beta}}{\partial s}
  &\frac{\partial \check{\beta}}{\partial n}
  &\frac{\partial \check{\beta}}{\partial v_s}
  &\frac{\partial \check{\beta}}{\partial v_n}
  &\frac{d^2}{\check{d}^2}
  \end{bmatrix},
\quad
  \frac{\partial{\mathbf{h}}}{\partial{\mathcal{X}}} = 
  \begin{bmatrix}
  \frac{\partial \theta}{\partial s} &\frac{\partial \theta}{\partial n} &0  &0       &0\\
  \frac{\partial \varphi}{\partial s}         &\frac{\partial \varphi}{\partial n}&0 &0 &0\\
  \frac{\partial \mu}{\partial s} &\frac{\partial \mu}{\partial n} &\frac{\partial \mu}{\partial v_s}         &\frac{\partial \mu}{\partial v_n}       &0\\
  \frac{\partial \tau}{\partial s} &\frac{\partial \tau}{\partial n} &0         &0       &0\\
  0 &0 &0         &0       &1\\
  \end{bmatrix}.
\end{equation}
where we omit the time index $l$-$1$ at the $l$-th epoch and substitute $(\cdot)_l$ as $\check{(\cdot)}$.
The derivates in the two matrices can be calculated by recalling the definition of the state evolution model and the measurement model and applying the chain rule.

We are now ready to present the EKF technique.
By applying the standard procedure of Kalman filtering\cite{kay1993fundamentals}, the state prediction and tracking design can be summarized as follows:

\noindent 1) \emph{State Prediction:}\label{State-Prediction}
\begin{equation}
  \hat{\mathcal{X}}_{l|l-1} = \mathbf{g}(\hat{\mathcal{X}}_{l-1}),
\end{equation}
2) \emph{Linearization:}
\begin{equation}\label{Linearization}
  \mathbf{G}_{l-1} = \left.\frac{\partial{\mathbf{g}}}{\partial{\mathcal{X}}}\right|_{\mathcal{X} = \hat{\mathcal{X}}_{l-1}},
  \mathbf{H}_{l}   = \left.\frac{\partial{\mathbf{h}}}{\partial{\mathcal{X}}}\right|_{\mathcal{X} = \hat{\mathcal{X}}_{l|l-1}},
\end{equation}
3) \emph{MSE Matrix Prediction:}
\begin{equation}\label{MSE-Matrix-Prediction}
  \mathbf{M}_{l|l-1} = \mathbf{G}_{l-1}\mathbf{M}_{l-1}\mathbf{G}^H_{l-1}+\mathbf{Q}_s,
\end{equation}
4) \emph{Residual Covariance and Kalman Gain Calculation:}
\begin{align}\label{Kalman-Gain-Calculation}
  \mathbf{S}_l      = \mathbf{H}_{l} \mathbf{M}_{l|l-1} \mathbf{H}^H_{l}+\mathbf{Q}_m,\quad
  \mathbf{K}_{l}    = \mathbf{M}_{l|l-1}\mathbf{H}_{l}^H\mathbf{S}_l^{-1},
\end{align}
5) \emph{Measurement Residual and State Update:}
\begin{align}\label{State-Update}
  \tilde{\mathcal{Y}_l}   = \mathcal{Y}_{l}-\mathbf{h}(\hat{\mathcal{X}}_{l|l-1}),\quad
  \hat{\mathcal{X}}_l     = \hat{\mathcal{X}}_{l|l-1} + \mathbf{K}_{l}\tilde{\mathcal{Y}_l},
\end{align}
6) \emph{MSE Matrix Update:}
\begin{equation}\label{MSE-Matrix-Update}
  \mathbf{M}_l = (\mathbf{I} - \mathbf{K}_l\mathbf{H}_l)\mathbf{M}_{l|l-1}.
\end{equation}

\subsection{IMM Filtering for Motion Tracking and Reasoning}
While the EKF scheme is capable of achieving high accuracy when tracking and predicting the state of the vehicle under a single kinematic model, it may be difficult to generalize to the case where the vehicle may switch among multiple kinematic models. 
This implies that the motion of the target may not be matched with the predicted motion of the given tracking model when the vehicles are maneuvered by the drivers.
For example, the lateral velocity is considered noise when the state is predicted by an LK model while the vehicle is changing the lane.
This can lead to severe prediction error and even loss of tracking.
Adding some additional state variables which describe the maneuver in other dimensions may help, such as applying the LC model.
However, the system noise may also be introduced when the models with more state variables are considered, which leads to performance degeneration when the vehicle is stably moving.
To improve the robustness of the tracking when maneuver occurs and guarantee the effectiveness when the vehicle is in a stable state, we invoke the IMM filtering algorithm.

\begin{figure}[!htbp]
  \centering
  \includegraphics[width=0.45\textwidth]{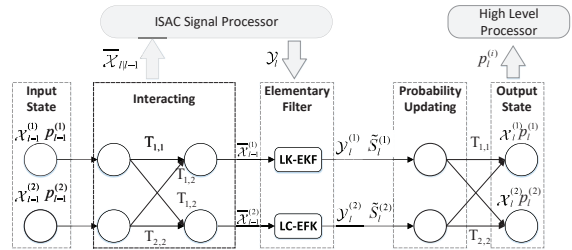}
  \caption{A flowchart of the IMM filter scheme.}\label{IMM_Demo}
  \end{figure}

An IMM filter scheme is designed for estimating the state of a system that may be matched with more than one potential model (hypothesis).
As shown in Fig. \ref{IMM_Demo}, an IMM filtering scheme consists of three steps: interacting, elementary filtering and model probability updating and combining.
The estimation and prediction of each model are calculated by a Kalman filter in the second step, and the results of different models interact at the initial and the final step of an IMM filter.
Compared with Kalman filters with only one hypothesis, the transition probability matrix $\mathbf{T}$ between different hypotheses is needed as prior information.
The entry in the $j$-th column and the $i$-th row of $\mathbf{T}$ represents the transition probability from the $j$-th model to the $i$-th model, which is irrelative to the estimation but only depends on the kinematic model and road geometry.

In the first step of IMM filter scheme at the $l$-th epoch, 
the probability, estimated state, MSE matrix and predicted state are regarded as input.
We firstly calculate the mixed probability, which represents the weight of each model at this epoch.
By mixing the estimated probability of each model and the transition probability matrix, the mixed probability can be expressed as
\begin{equation}
  c ^{(i|j)}_{l}       =\frac{T_{j,i} p_{l-1}^{(j)}}   {\sum\limits_{j}T_{j,i} p_{l-1}^{(j)}},\\
\end{equation}
where $p_{l-1}^{(j)}$ is the estimated probability of the $j$-th model at the ($l$-1)-th epoch.
Then, the input of the $i$-th elementary filter can be calculated by mixing the output of all the filters and the mixed probability $c$ as follow
\begin{equation}\label{Mixed-X}
  \bar{\mathcal{X}}^{(i)}_{l-1}  = \sum\limits_{j}\hat{\mathcal{X}}_{l-1}^{(j)} c_l^{(i|j)},
\quad
  \bar{\mathcal{X}}^{(i)}_{l|l-1}  = \sum\limits_{j}\mathbf{h}^{(i)}(\hat{\mathcal{X}}_{l-1}^{(j)}) c_l^{(i|j)},
\end{equation} 
\begin{align}
  \bar{M}^{(i)}_{l-1} =\sum\limits_{j} c^{(i|j)}_{l} \times
  (M^{(j)}_{l-1} + 
  (\bar{\mathcal{X}}^{(i)}_{l-1} -\hat{\mathcal{X}}_{l-1}^{(j)}) 
  (\bar{\mathcal{X}}^{(i)}_{l-1} -\hat{\mathcal{X}}_{l-1}^{(j)}) ^T).\label{Mixed-M}
\end{align}
Besides, to construct the transmit beamformer, it is necessary to calculate the mixed prediction for the interacted model by the following equation,
\begin{equation}
  \bar{\mathcal{X}}_{l|l-1} = \sum \limits_{i}\bar{\mathcal{X}}^{(i)}_{l|l-1}/N,
\end{equation}
where $N$ is the number of hypotheses.
\footnote{Although we regard the interacting step as the first step of the IMM filter, we need to calculate all these variables at the end of the previous epoch to output the best prediction of the best beam pattern in this epoch in (\ref{beampattern}).}

In the second step, the standard EKF procedures introduced in section \ref{EKF} are employed as elementary filters. By substituting the input variables for the single-model-filter as their interacted counterpart for multiple models in (\ref{Mixed-X})-(\ref{Mixed-M}), the elementary filtering can be easily realized followed by the procedure in Sec. \ref{EKF}.
Importantly, compared with a standard EKF scheme for a single model, the output of an elementary filter not only contains the estimated state, the predicted state and the MSE matrix, but also includes the measurement residual and the residual covariance for each element.
For clarity, we rewrite them in the interacted form for the $i$-th element as
\begin{equation}
  \tilde{\mathcal{Y}}_l^{(i)} = \mathcal{Y}_{l}-\mathbf{h}^{(i)}(\bar{\mathcal{X}}^{(i)}_{l|l-1}),
\quad
  \tilde{S}^{(i)}_l = \mathbf{H}^{(i)}_{l} \mathbf{M}^{(i)}_{l|l-1} \mathbf{H}^{(i) \, H}_{l}+\mathbf{Q}_m.
\end{equation}

In the last step, the likelihood $L$ and the probability $p$ for the $i$-th model at the $l$-th epoch are updated by analyzing the measurement residual and the residual covariance matrix by the following equations:
\begin{align}
    L_{l}^{(i)} = \frac{\exp\left( {-\frac{1}{2}     \left( \tilde{\mathcal{Y}}_l^{(i)}\right)^T  \left( \tilde{S}^{(i)}_l \right)^{-1}  \tilde{\mathcal{Y}}_l^{(i)}} \right)}{(\sqrt{(2\pi)^{n}|\tilde{S}^{(i)}_l|}},\quad
    p^{(i)}_l = \frac{L_{l}^{(i)} \sum\limits_{j}T_{j,i} p_{l-1}^{(j)} }{ \sum\limits_i \left( L_{l}^{(i)} \sum\limits_{j}T_{j,i} p_{l-1}^{(j)} \right)}.
\end{align}
By following the steps above, the fitness between the actual state and the hypothesis is measured by a probability, and the beamformer is almost determined by the model with the best matching probability..
Besides, the overall output state and corresponding MSE matrix can be express as
\begin{align}
  \hat{\mathcal{X}}_l &= \sum \limits_{j} \hat{\mathcal{X}}_l^{(j)}  p_l^{(j)},\\
  M_{l} &= \sum\limits_{j} p^{(j)}_{l} \times (M^{(j)}_{l} + 
  (\hat{\mathcal{X}}^{(i)}_{l} -\hat{\mathcal{X}}_{l}) 
  (\hat{\mathcal{X}}^{(i)}_{l} -\hat{\mathcal{X}}_{l}) ^T),
\end{align}
which can be employed to analyze the estimation and prediction  performances.

\section{Dynamic Beamwidth Scheme for ISAC V2I Link}\label{DB}
The transmit beam should be as narrow as possible to achieve high array gain once the accurate location of the vehicle is predicted.
However, the accuracy of the IMM-EKF algorithm is limited by both the system noise and measurement noise. This indicates that under the massive-MIMO assumption, the vehicle may locate at the sidelobe and the beam will point to a wrong direction, leading to a low array gain.
In this section, we analyze the uncertainty of the location and propose a dynamic beamwidth scheme under a given misalignment probability requirement.

\subsection{Predicted Uncertainty for Vehicles}
To characterize the uncertainty of the prediction, we derive the distribution of the predicted state which has a direct impact on beamforming.
Following the definition of the MSE matrix in the EKF algorithm, the expectation of error between the predicted state and the true state can be given as
\begin{align}
   &\mathbb{E}\{(\hat{\mathcal{X}}_{l|l-1} -\mathcal{X}_l) (\hat{\mathcal{X}}_{l|l-1} -\mathcal{X}_l)^H\} \notag\\
  =&\text{E}\{
    (g(\hat{\mathcal{X}}_{l-1})\!-\!g(\mathcal{X}_{l-1})\!-\!\bm{\omega}_l)
    (g(\hat{\mathcal{X}}_{l-1})\!-\!g(\mathcal{X}_{l-1})\!-\!\bm{\omega}_l)^H
    \} \notag\\ 
  \overset{(a)}{\approx}& \text{E}\{
    \big(\mathbf{G}_{l-1}(\hat{\mathcal{X}}_{l-1}\!-\!\mathcal{X}_{l-1})\!-\!\bm{\omega}_l\big)
    \big(\mathbf{G}_{l-1}(\hat{\mathcal{X}}_{l-1}\!-\!\mathcal{X}_{l-1})\!-\!\bm{\omega}_l\big)^H
    \} \notag \\
  \overset{(b)}{=}& \mathbf{G}_{l-1} \mathbf{M}_{l-1} \mathbf{G}_{l-1}^H + \mathbf{Q}_s = \mathbf{M}_{l|l-1},
\end{align}
where the linear approximation $(a)$ follows by the linearization step of EKF and $(b)$ follows by the fact that the system noise $\bm{\omega}_l$ is independent to $\mathcal{X}_{l-1}$ and $\hat{\mathcal{X}}_{l-1}$, i.e., $\text{E}\{\mathcal{X}_{l-1} \bm{\omega}_l^H\} = 0$ and $\text{E}\{\hat{\mathcal{X}}_{l-1} \bm{\omega}_l^H\} = 0$.
Since both $\hat{\mathcal{X}}_{l-1}$ and $\bm{\omega}_l$ are Gaussian distributed, $\hat{\mathcal{X}}_{l-1}$ is Gaussian distributed as well, which subject to
\begin{equation}
  \hat{\mathcal{X}}_{l|l-1} \sim \mathcal{CN}(\mathcal{X}_l,\mathbf{M}_{l|l-1}).
\end{equation}

As our main purpose is to find the proper beamwidth, we focus on the state variables $s$ and $n$, which are relative to $\varphi$ and $\theta$.
The uncertainty of the angle can be modeled as
\begin{equation}
  \Psi 
  \!=\!
  \left[\begin{matrix}
    \Delta \theta_l \\ \Delta \varphi_l
  \end{matrix}\right]
  \!=\!
  \left[\begin{matrix}
    \hat{\theta}_{l|l-1} - \theta_l\\ \hat{\varphi}_{l|l-1} - \varphi_l
  \end{matrix}\right]
  \!\approx \!
  \left[\begin{matrix}
    \frac{\partial \theta}{\partial s}   \! & \! \frac{\partial \theta}{\partial n}\\
    \frac{\partial \varphi}{\partial s}  \! & \! \frac{\partial \varphi}{\partial n}
    \end{matrix}\right]
  \left[\begin{matrix}
    \hat{s}_{l|l-1} - s_l \\ \hat{n}_{l|l-1} - n_l
  \end{matrix}\right],
\end{equation}
where we apply the first order linear approximation in the above equation due to the non-linearity between the angle and the state variables.
With the distribution of predicted state $\hat{\mathcal{X}}_{l|l-1}$ in hand, one can directly express the distribution of $\Psi$ as
\begin{equation}\label{Calculate-Sigma}
  \Psi \sim \mathcal{N}(0,\Sigma),\quad \Sigma = 
  \left[\begin{matrix}
    \frac{\partial \theta}{\partial s}   & \frac{\partial \theta}{\partial n}\\
    \frac{\partial \varphi}{\partial s}   & \frac{\partial \varphi}{\partial n}
    \end{matrix}\right]
    \mathbf{M}_{l|l-1}
    \left[\begin{matrix}
      \frac{\partial \theta}{\partial s}   & \frac{\partial \theta}{\partial n}\\
      \frac{\partial \varphi}{\partial s}   & \frac{\partial \varphi}{\partial n}
      \end{matrix}\right]^T.
\end{equation}

\subsection{Problem Formulation}
In the massive MIMO regime, the array gain decreases drastically when the target deviates from the center of the transmit beam. Thus, we define the area where the normalized array gain is no less than -3dB of its peak as \emph{3dB-beamwidth} and the probability of the target out of this area as \emph{misalignment probability}.
Following the definition above, we firstly unveil the relationship between the array gain and $\Psi$  for any given size of the antenna array.
By recalling (\ref{TxGain}), the array gain with respect to the amizuth angle and elevation angle can be derived from beamformer and channel matrix in the following form 
\begin{align}\label{AG-Delta}
  &\kappa(\theta_l,\varphi_l,\hat{\theta}_{l|l-1},\hat{\varphi}_{l_l-1})\! =\\ \notag &\big|\sum\limits_{n=0}^{N-1} \sum\limits_{m=0}^{M-1} \exp\{ j\pi \Big(n(\cos\theta_l\sin\varphi_l\!-\!\cos\hat{\theta}_{l|l-1}\sin\hat{\varphi}_{l_l-1}) \!+\! m(\sin\theta_l\sin\varphi_l\!-\!\sin\hat{\theta}_{l|l-1}\sin\hat{\varphi}_{l_l-1})\big)\}\Big|,
\end{align}
To give a more compact form to find the 3dB region, we approximate the array gain as
\begin{equation}
  \kappa^2 \! \approx \! \Big| \! \sum\limits_{n=0}^{N-1} \!\sum\limits_{m=0}^{M-1} \! e^ {j\pi(n \Psi_{\parallel } +  m\Psi_{\bot})}\Big|^2, 
  \label{Antenna-gain}
\end{equation}
with  $\Psi_{\parallel } = -\sin\theta\sin\varphi \Delta \theta + \cos \theta \cos \varphi \Delta \varphi $ and $\Psi_{\bot } = \sin\theta\cos\varphi \Delta \theta + \cos \theta \sin \varphi \Delta \varphi $, and time indexes are omitted in the rest of this section.\footnote{While the real value of the angle is unavailable, we use the predicted value as a approximated counterpart.The performance loss of the approximation is neglectable when the tracking and prediction has a high precision.}
So, the distribution of $\tilde{\Psi}$ can be expressed as
\begin{align}
  \tilde{\Psi} \!=\! [\Psi_{\parallel},\Psi_{\bot}]^T \!\sim\! \mathcal{N}(0,\tilde{\Sigma}), \tilde{\Sigma} = \mathbf{T}_{\Sigma}  \Sigma  \mathbf{T}_{\Sigma}^T,
  \label{Calculate-tilde-Sigma}
\end{align}
with $\mathbf{T_{\Sigma}}$ being the transition matrix from the angles to spatial frequencies, which can be mathematically expressed as
\begin{equation}
  \mathbf{T_{\Sigma}} = 
  \left[\begin{matrix}
    -\sin\theta\sin\varphi & \cos \theta \cos \varphi\\
     \sin\theta\cos\varphi & \cos \theta \sin \varphi 
  \end{matrix}\right].
\end{equation}
The corresponding PDF can be written as
\begin{equation}
  f(\tilde{\Psi}) = \frac{1}{2\pi|\tilde{\Sigma}|^{\frac{1}{2}}}e^{-\frac{1}{2} \tilde{\Psi} \tilde{\Sigma}^{-1} \tilde{\Psi}^T}.
  \label{PSI-PDF}
\end{equation}

Since the 3dB-beamwidth and the misalignment probability for any size of the antenna array can be numerically calculated by (\ref{Antenna-gain}) and (\ref{PSI-PDF}), one may check whether the misalignment probability exceeds the threshold and find the minimum beamwidth, with the highest array gain, by simply employing a two-dimensional linear search method, which could be computationally expensive.
Thus, we propose an algorithm that finds a proper region $D_0$ covered by 3dB-beamwidth with a minimum area in which the cumulative probability of the vehicle is no less than $1-\Gamma$.
This task can be mathematically expressed as
\begin{align}
  &\min \quad D_0  \notag \\
  &s.t. \quad \iint \limits_{D_0} f(\tilde{\Psi})dD > 1-\Gamma.
\end{align}
Nevertheless, the region is an eclipse on the $\theta$-$\varphi$ plane, on which the double integral of $f(\tilde{\Psi})$ is difficult to be expressed in a closed-form.
We resort to integrating the PDF along its natrual axis.
By defining the radius sacale $r$ as $r^2 =  \tilde{\Psi} \tilde{\Sigma}^{-1} \tilde{\Psi}^H$, the cumulative probability in the area where $r<R_0$ can be expressed as
\begin{align}
  F(R_0) &= \iint \limits_{D_1(R_0)} f(\tilde{\Psi}) dD = \int \limits_{0}^{R0} f(r) dD \notag \\ 
         &= \int \limits_{0}^{R_0} \frac{1}{2\pi|\tilde{\Sigma}|^\frac{1}{2}}e^{-\frac{1}{2} r^2} 2\pi|\tilde{\Sigma}|^{\frac{1}{2}}r dr = 1-e^{-\frac{1}{2} R_0^2}.\label{Radius-scale}
\end{align}
Accordingly, for any given $\Gamma$, it is possible to find a corresponding $R_0$ which guarantees the cumulative probability being equal to $1-\Gamma$.
\begin{figure}
  \centering
  \includegraphics[width=0.45\textwidth]{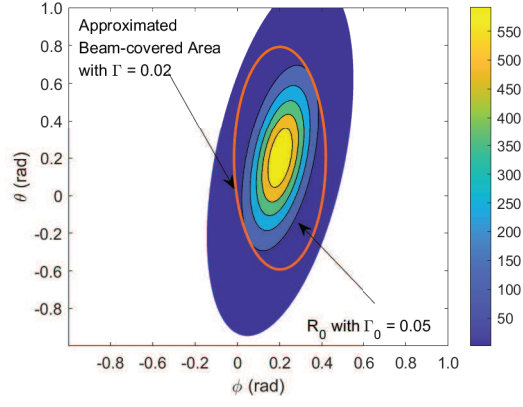}
  \caption{Contour plot of $f(\tilde{\Psi})$, best area with $\Gamma_0$ and the approximated beam-covered area.}\label{DB_Demo}
\end{figure}

Since the covariance matrix $\tilde{\Sigma}$ cannot be approximated as a diagonal matrix, the area $D_1$ is a rotated eclipse, which indicates that we can not directly use the calculated $R_0$ to obtain the optimal beamwidth.
Therefore, as shown in Fig. \ref{DB_Demo}, we turn to find the sub-optimal region $\tilde{D}_0 \supseteq  D_1(R_0) $ where the misalignment probability is less than $\Gamma$ which results in little performance loss.
By simple geometric derivation, the expression of the sub-optimal 3dB-beamwidth which covers $D_1(R_0)$ with highest array gain can be given as
\begin{equation}
  \tilde{D}_0:   A_{\theta}\theta^2 +B_{\varphi}\varphi^2 \leq 1.\label{sub-optimal-area}
 \end{equation}
In the above equation, $A_{\theta}$ and $B_{\varphi}$ are the parameters of the semi-major axis and semi-minor axis of $D_0$, which can be expressed as
\begin{align}
 A_{\theta} =  \Sigma^{-1}_{1,1} - \|{\Sigma}^{-1}_{2,1}|\sqrt{\Sigma^{-1}_{1,1} /{\Sigma^{-1}_{2,2}}  }/2,\quad
B_{\varphi} =  \Sigma^{-1}_{2,2} - \|{\Sigma}^{-1}_{1,2}|\sqrt{\Sigma^{-1}_{2,2} /{\Sigma^{-1}_{1,1}}  }/2,
\label{Axis-Eclipse}
\end{align}
with ${\Sigma}^{-1}_{i,j}$ being the elment in $i$-th colomn and $j$-th row of the inversed covariance matrix.

Finnaly, the relationship between the size of choosen subarray and the beamwidth can be expressed as
\begin{align}
  \theta_{BW} = \sqrt{1/A_{\theta}} =\frac{k_0 \lambda}{N} \approx \frac{0.89}{N},\quad
  \varphi_{BW}= \sqrt{1/B_{\varphi}}   =\frac{k_0 \lambda}{M} \approx \frac{0.89}{M},
\end{align}
where $k_0$ denote the bandwidth factor.
The best antenna size of the antenna array can be directly given as
\begin{equation}
  N^*= \left\lfloor 0.89 \sqrt{\frac{\tilde{\Sigma}_{2,1}\sqrt{\tilde{\Sigma}_{1,1}}}{8\sqrt{\tilde{\Sigma}_{2,2}}}}\right\rfloor,
  M^* = \left\lfloor 0.89 \sqrt{\frac{\tilde{\Sigma}_{2,1}\sqrt{\tilde{\Sigma}_{2,2}}}{8\sqrt{\tilde{\Sigma}_{1,1}}}}\right\rfloor.\label{best-size} 
\end{equation}
The overall algorithm is summarized in Algorithm 1.

\renewcommand{\algorithmicrequire}{\textbf{Input}}
\renewcommand{\algorithmicensure}{\textbf{Output}}
\begin{algorithm}
\caption{Dynamic Beamwidth scheme}
\label{alg1}
\begin{algorithmic}
    \REQUIRE :
             Desired misalignment probability $\Gamma$, 
             predicted State $\hat{\mathcal{X}}_{l|l-1}$,
             predicted MSE matrix $\hat{\mathbf{M}}_{l|l-1}$,
             Jacobian matrix for evolution function $G_{l-1}$.
    
    \ENSURE The best size of antenna array $M^{*}$ and $N^{*}$.
    \STATE 1. Calculate the covariance matrix by using (\ref{Calculate-Sigma}) and (\ref{Calculate-tilde-Sigma}) .
    \STATE 2. Get the radius scale factor $R_0$ by calculating $F(R_0)<1-\Gamma$ in (\ref{Radius-scale}).
    \STATE 3. Normalize the covariance matrix and calculate the corresponding sub-optimal beam-covered area by using (\ref{sub-optimal-area}).
    \STATE 4. Calculate the best size of antenna array in (\ref{best-size}).
\end{algorithmic}
\end{algorithm}

\section{Numerical Results}
\begin{figure}[!htbp]
  \begin{minipage}[t]{0.48\textwidth}
  \centering
  \subfigure [Country road]{
  \includegraphics [width=0.45\textwidth]{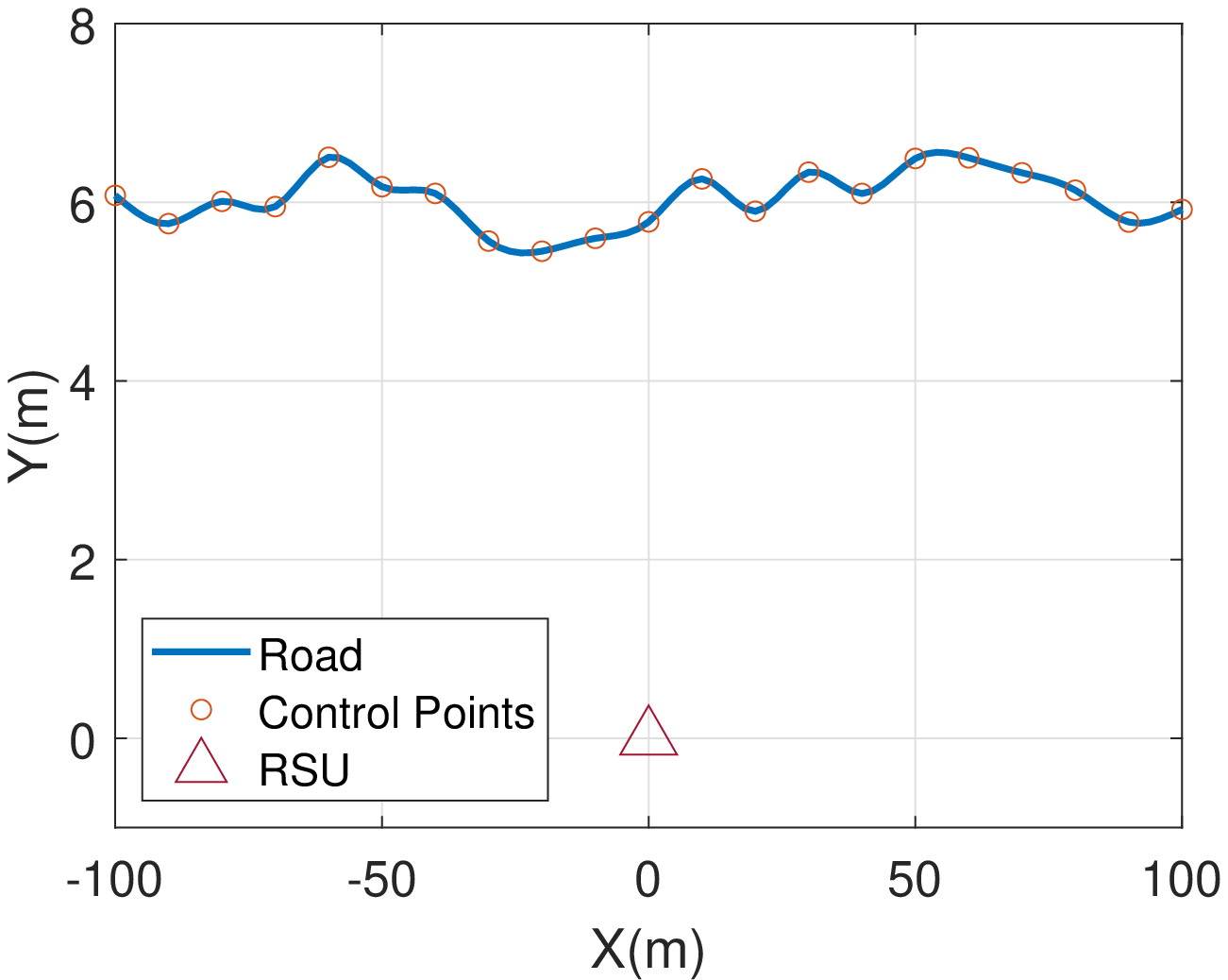}}
  \subfigure [Roundabout]{
  \includegraphics [width=0.45\textwidth]{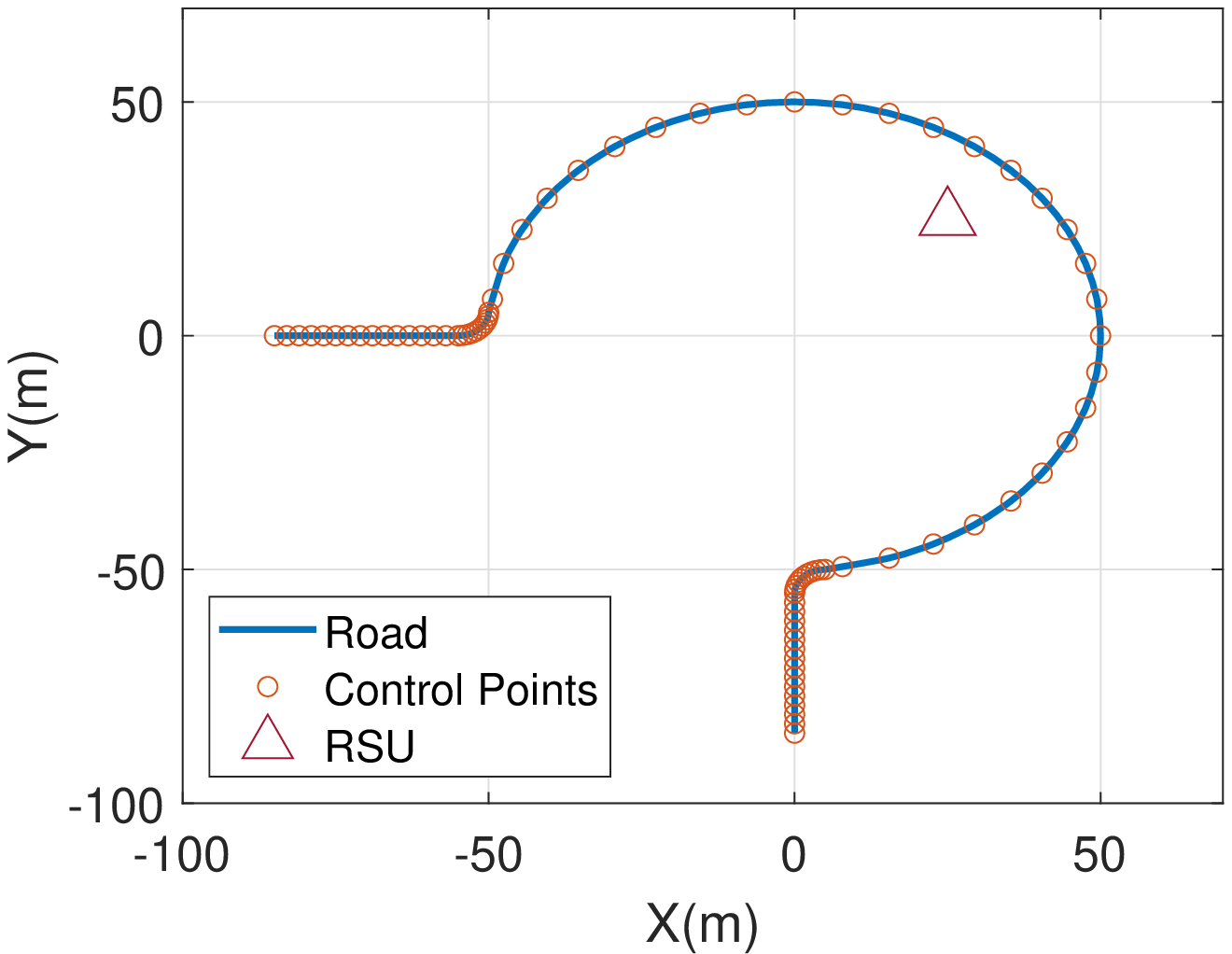}}
  \caption{Two road models used in simulation.}\label{RoadModel}
  \end{minipage}
  \quad
  \begin{minipage}[t]{0.48\textwidth}
    \centering
    \subfigure [Country Road]{
    \includegraphics [width=0.45\textwidth]{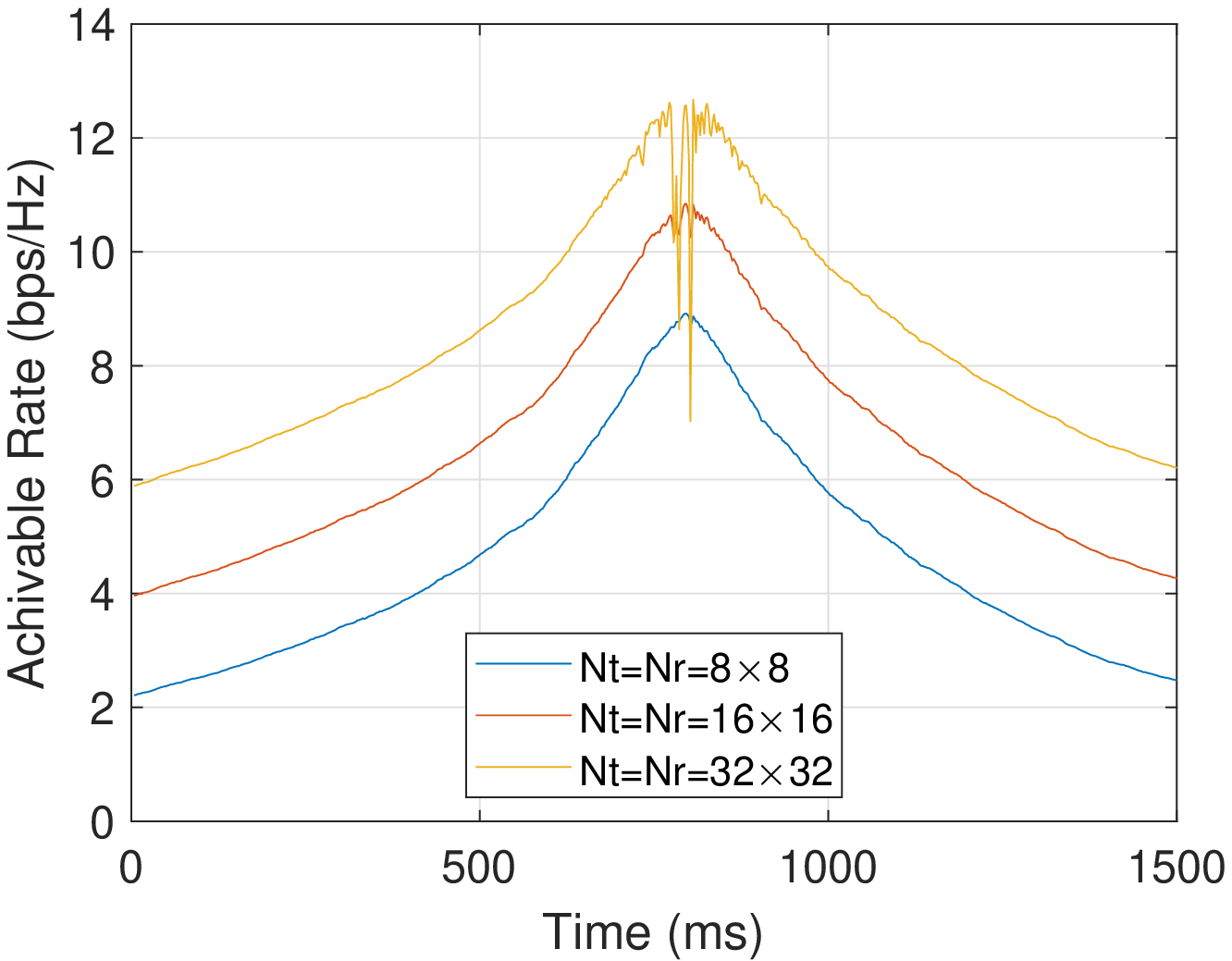}}
    \subfigure [Roundabout]{
    \includegraphics [width=0.415\textwidth]{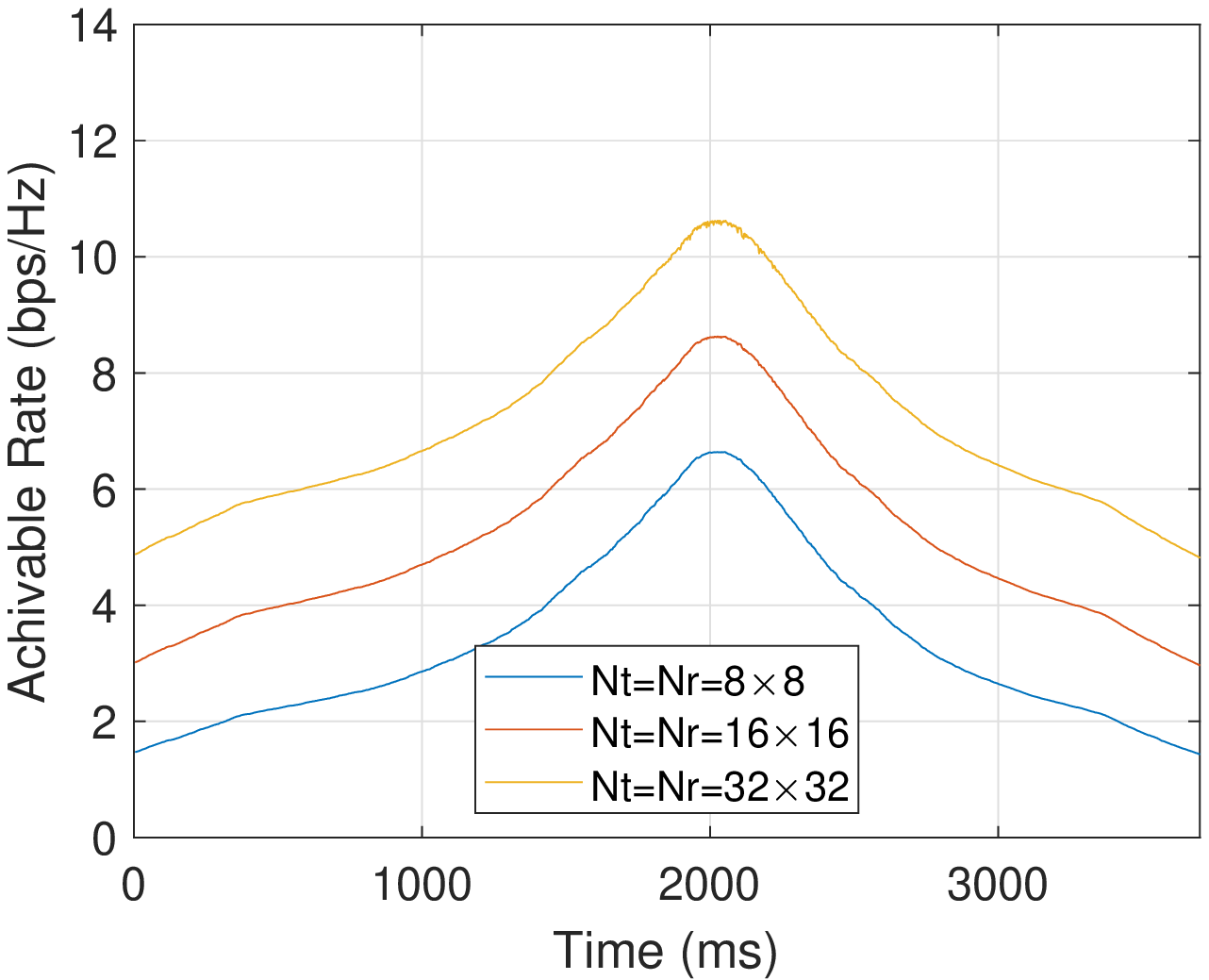}}
    \caption{Achievable rate for proposed scheme, with initial state $s = 2$m, $v_s = 10$m/s, $n = 0$m, $\beta = 2 \times 10^{-5} \times( 1+j)$, $P_t = 30$dBm.}\label{EKF_RG_ACR}
  \end{minipage}
 \end{figure}

In this section, we evaluate the performance of our proposed scheme in both sensing and communication functionalities by using numerical simulation.
Unless otherwise specified, both the RSU and the vehicle operate at $f_c$ = 30GHz with a bandwidth $B_w$ as 500MHz, the transmit power $P_t$ is considered as $P_t=30$dBm, and the data-frame block duration is set as $\Delta T$ = 20ms. 
The noise power spectral density is considered as $-144$ dBm/Hz at all the receivers, which leads to the variance of the noise as $-57$ dBm. The path losses for both sensing and communication functions are calculated by applying the free space transformation model.
The road models applied in this section are shown as Fig.\ref{RoadModel}.
In the first model, a country road is considered where the coverage of the RSU is set as 200m and the minimum distance from the road to RSU is set as 6m.
In the second model, the radius of the roundabout is set as 50m, which follows existing standards in China where the maximum velocity of the vehicle is limited to 10m/s\cite{UrbanRoad}, and the origin of the Cartesian coordinate system is set as the center of the roundabout.
The RSU is located at $\{25,25\}$ in the Cartesian system to obtain the ability to sense the Doppler shift along the road.
The variances of system noise for both model are set as follow:
$\sigma_{s}$= 0.08m, $\sigma_{n}$= 0.016m, $\sigma_{v_s}$= 0.1m/s, $\sigma_{v_n}$ = 0.02m/s, $\sigma_\beta = 2 \times 10^{-6}$, and the covariance matrix of the measurement noise is calculated in EKF procedure by applying (\ref{CRLB_angle}) and (\ref{CRLB_beta}-\ref{CRLB_mu}).
Moreover, to separately explore the improvement of each proposed scheme, the dynamic beamwidth scheme is only applied in the last subsection.

\

\subsection{Performance for Tracking the Vehicle in the LK Model}
We first evaluate the communication and sensing performances of the proposed technique under the LK assumption. Without loss of generality, we set the initial state of the vehicle as $s = 2$m, $v_s = 10$m/s, $n = 0$m, $\beta = 2 \times 10^{-5} \times( 1+j)$.

In Fig. \ref{EKF_RG_ACR}, we show the downlink achievable rates of the communication link with increased size of the antenna array.
In both scenarios, the initial location of the vehicle is far from the RSU, which leads to low SNRs and corresponding low achievable rates.
When the vehicle is approaching the RSU, the path-losses decrease with respect to the distance, which accordingly leads to increasing achievable rates.
Note that, the communication performance degenerates when the vehicle is quite near to the RSU in the country road model when $N_t$ = 32 $\times$ 32.
On the one hand, the velocity of the vehicle is almost vertical to the propagation direction of the signal, which makes the Doppler frequency hard to be estimated. On the other hand, since the distance between the RSU and the vehicle is sufficiently small, the angle changes too fast for EKF tracking.
Fortunately, this degeneration does not occur when the size of the antenna array is small and can also be compensated by the dynamic beamforming algorithm proposed in Sec. \ref{DB} when the size of the antenna array is large.

\begin{figure}[!htbp]
  \centering
  \subfigure [Longitudinal distance in country road.]{
  \includegraphics [width=0.22\textwidth]{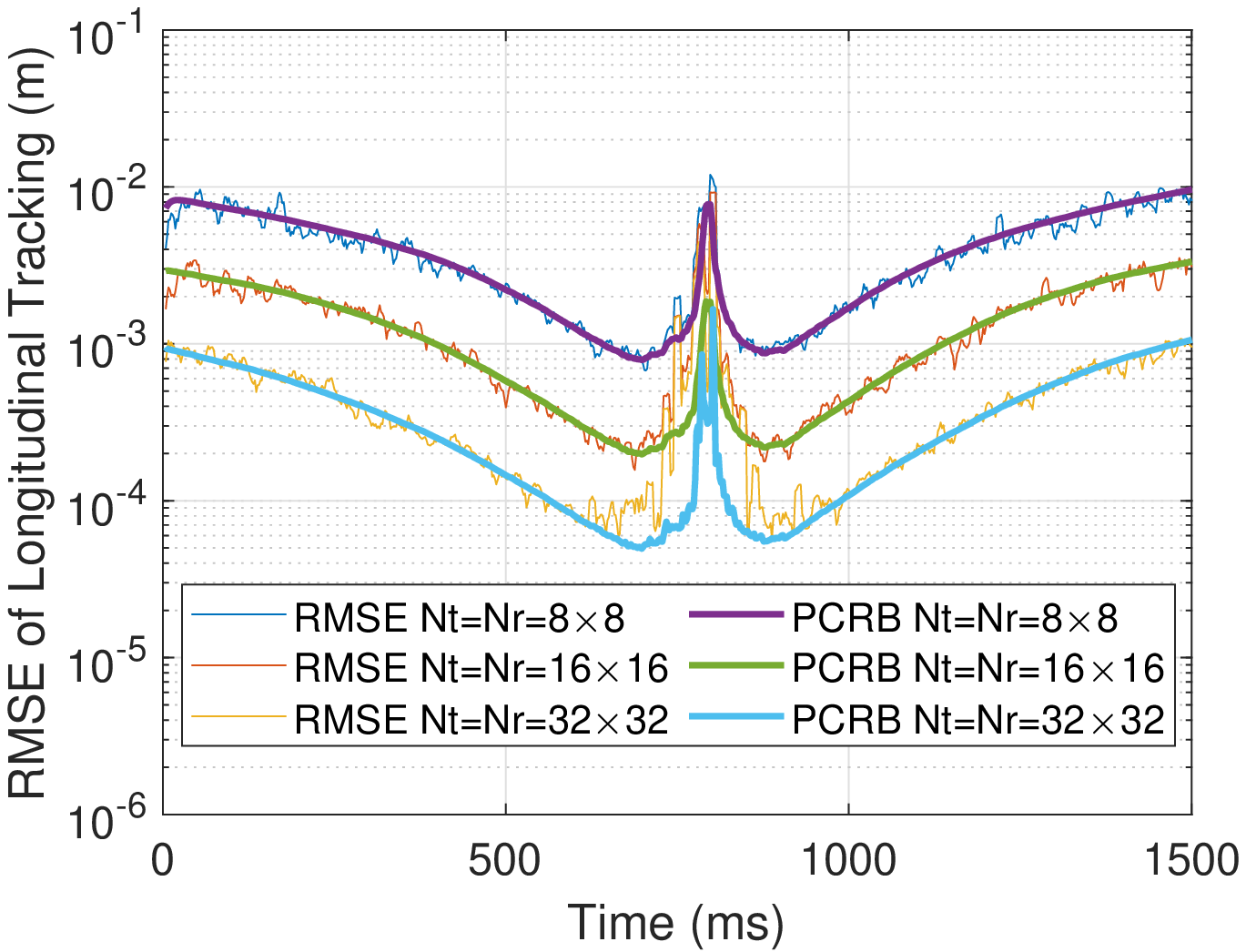}}
  \subfigure [Longitudinal distance in roundabout.]{
  \includegraphics [width=0.22\textwidth]{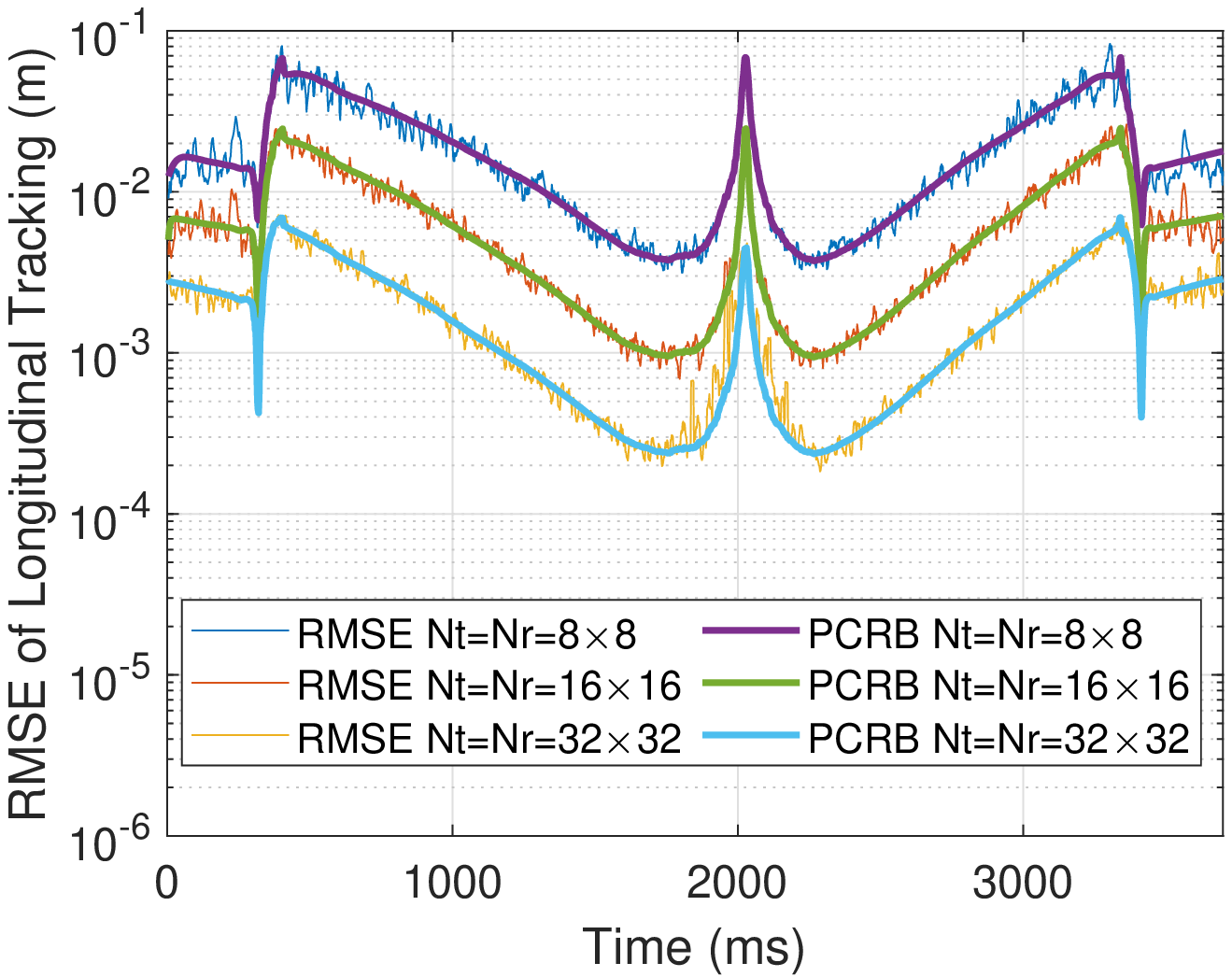}}
  \subfigure [lateral distance in country road.]{
  \includegraphics [width=0.22\textwidth]{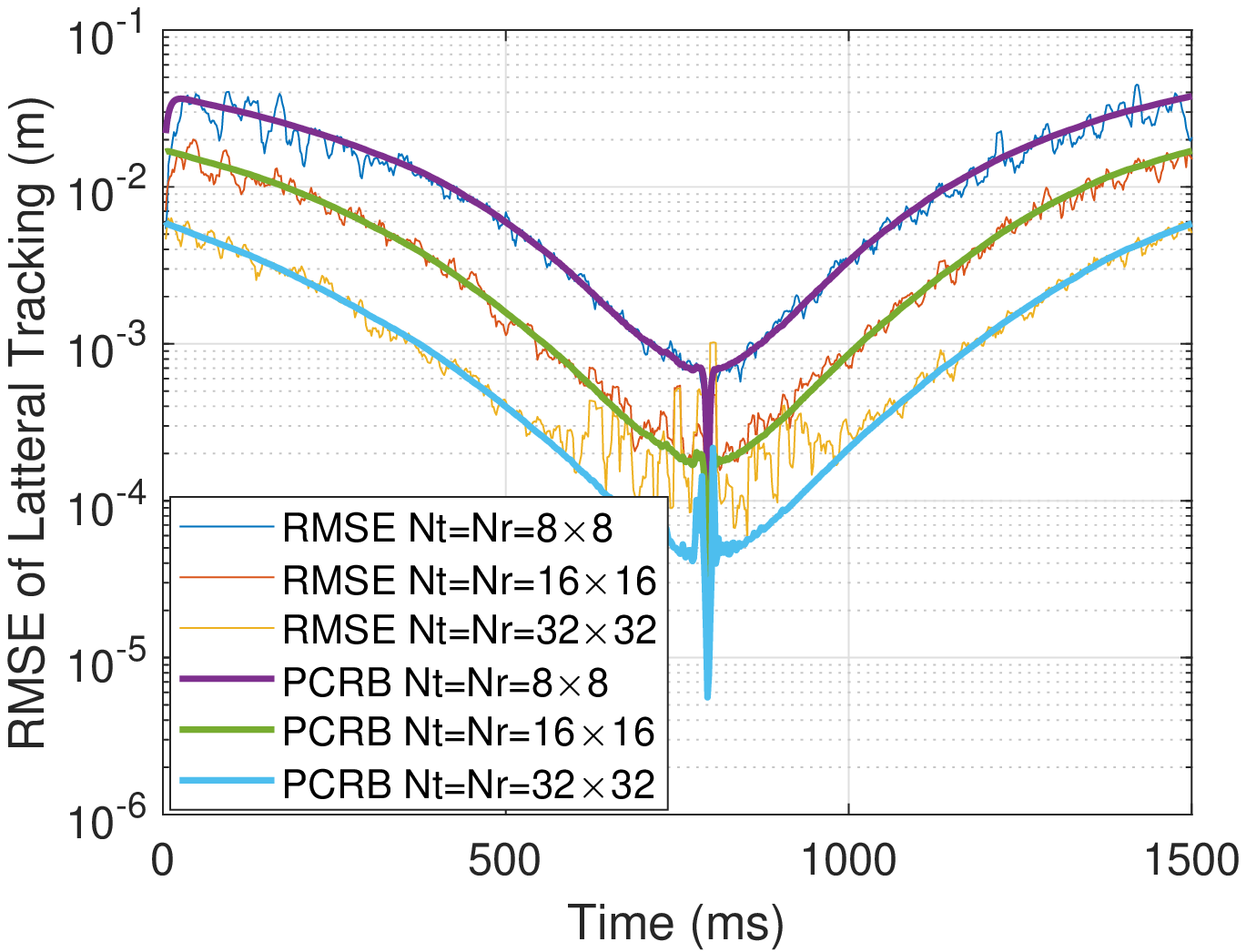}}
  \subfigure [lateral distance in roundabout.]{
  \includegraphics [width=0.22\textwidth]{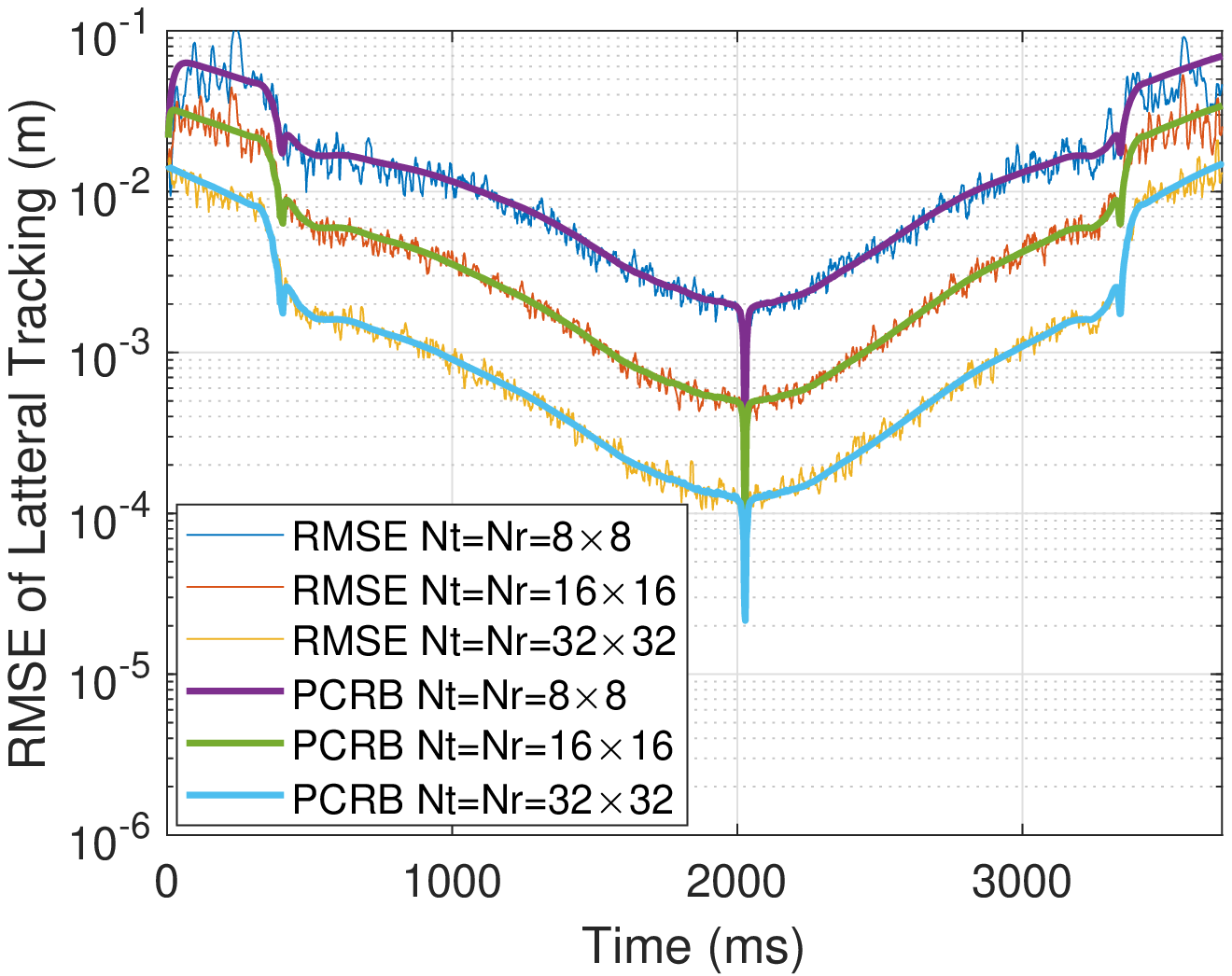}}  
  \caption{Position tracking performance for proposed scheme , with initial state $s = 2$m, $v_s = 10$m/s, $n = 0$m, $\beta = 2 \times 10^{-5} \times( 1+j)$, $P_t = 30$dBm.}\label{EKF_RG_Distance}
 \end{figure}

In Fig. \ref{EKF_RG_Distance}, we demonstrate the radar sensing performance in terms of root mean squared error (RMSE) for the position tracking performance in the CCS.
As shown in Fig. \ref{EKF_RG_Distance}, benefiting from the improvement of the SNR and the accuracy of the measurement in each epoch, the tracking performances generally increase when the vehicle is moving near the RSU.
Moreover, when the direction of the vehicle is almost perpendicular to the RSU (800ms in (a) and (c), and 2100ms in (b) and (d)), it is difficult to track the Doppler shift in the longitudinal direction while the tracking performance in the lateral direction is improved. 
The spikes in the RMSEs in (b) and (d) (at about 500ms and 3500ms) are due to the swift direction changes when entering and exiting the roundabout in Fig. 7(a).

It is noteworthy that, when the beam precisely points to the vehicle, the RMSE is tightly bounded by PCRB, which proves the effectiveness of the proposed scheme in Sec. \ref{DB} that dynamically adjusts the beamwidth by leveraging the predicted MSE matrix. 

\subsection{Performace Comparision for Proposed Algorithm and Benchmark Schemes}

In this subsection, we examine the superiorities of our proposed algorithm (ISAC-RG) by comparing it with the following beam tracking schemes in the country road model:
\begin{itemize}
  \item The ISAC-based beam tracking scheme in the Cartesian coordinate system, which predicts the state of the vehicle on the complicated road by applying a difference algorithm (ISAC-Cartesian).\cite{liu2020tutorial}.
  \item  The ISAC-based beam tracking scheme which assumes the vehicles are driven on a straight road (ISAC-SR).\cite{liu2020radar}.
  \item The auxiliary beam pair algorithm where quaternary training beams are transmitted to the vehicle at the beginning of an epoch (ABP)\cite{zhu2017auxiliary}.
\end{itemize}
Note that the ABP algorithm is a kind of communication-only beam tracking algorithm that needs additional uplink feedback to inform the RSU of the channel estimation, thus the overhead is much higher than that of ISAC-based algorithms.
The ISAC-Cartesian algorithm is applied to situations where the state evolution functions are difficult to derive and usually has some performance loss.
Here, the number of transmit and receive antennas is assumed to be 256, and the half searching range of both elevation and azimuth beamforming region for the ABP algorithm is set as ${\pi}/{8}$ without loss of generality.

\begin{figure}[!htbp]
  \begin{minipage}[t]{0.48\textwidth}
  \centering
  \includegraphics[width=1\textwidth]{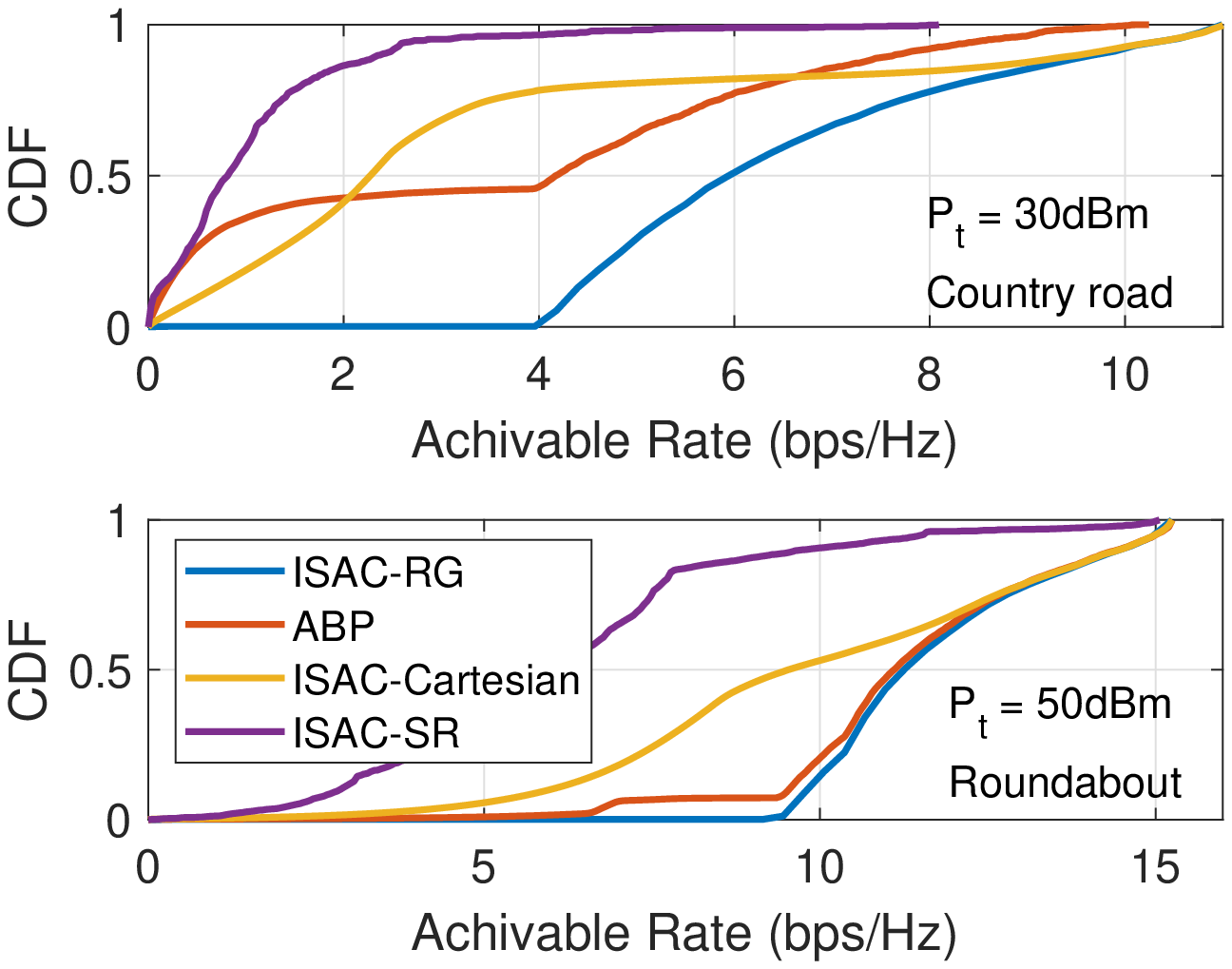}
  \caption{CDF of achievable rates for different alogrithms with $N_t = N_r = 256$, $P_t = 30$dBm and $50$dBm. }\label{Comparasion_ACR_CDF}
  \end{minipage}
  \quad
  \begin{minipage}[t]{0.48\textwidth}
    \centering
    \includegraphics[width=1\textwidth]{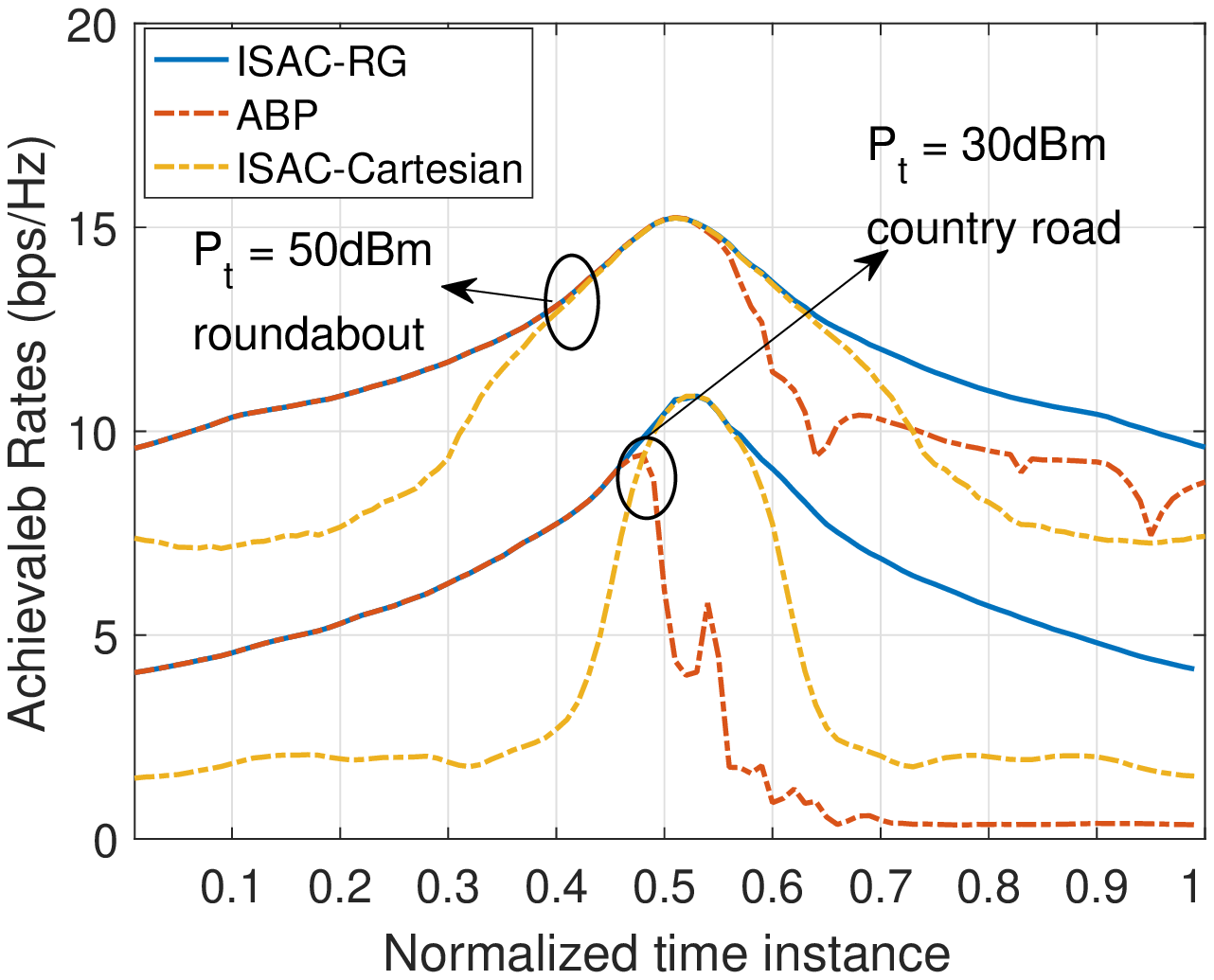}
    \caption{Achievable rates comparision for different alogrithms with $N_t = N_r = 256$, $P_t = 30$dBm and $50$dBm. }\label{Comparasion_ACR}   
  \end{minipage}  
\end{figure}

In Fig. \ref{Comparasion_ACR_CDF}, we firstly explore the communication performance by showing the empirical cumulative distribution functions (CDF) in terms of the achievable rates.
It is notable to see that the proposed algorithm always achieves the best performance compared with the benchmark schemes and shows robustness to the transmit power budget.
It can also be seen that the gap of communication performance between the ISAC-Cartesian and the proposed algorithm reduces as the transmit power increases, and the gap between the ABP and the ISAC-RG also decreases since the roundabout model is friendly to ABP algorithm.
As for the ISAC-SR, the mismatched model makes it difficult to track the vehicle properly and leads to severe performance loss.

We further show the achievable rates obtained for different schemes in one realization in Fig. \ref{Comparasion_ACR} to interpret how the movement of the vehicle affects the communication performance.
To be mentioned, we do not show the ISAC-SR in this figure since the reason for performance loss is obvious and the performance with a higher power budget has too many crossovers with that of other schemes with a lower budget.
It can be seen that the ABP algorithm shows comparable performance when the vehicle is moving to the RSU (0 to 0.5) and suffers severe performance loss after passing the RSU.
This is because the ABP algorithm uses the feedback direction to finish the beam tracking and the performance almost relies on the accuracy of angle estimation in the current epoch but not all the past epochs, which results in instability of tracking.
More specifically, once the angle begins to vary rapidly and the beamformer is slightly misaligned, a biased estimation may occur which results in the loss of the track.
Moreover, because the minimum distance from the vehicle to the RSU in the roundabout is much larger than that in the country road, the angular velocity is smaller and leads to fewer performance losses. 
As for the ISAC-Cartesian algorithm, since the prediction performance only relies on the measurement in the last three steps and the RMSE of the prediction is several times of the RMSE of the estimation, the achievable rate is more relevant to radar SNR compared with the proposed algorithm.
As shown in the figure, the performance is comparable with the proposed algorithm in the area near the RSU where the radar SNR is high enough, but severe performance degenerations occur in the other area. 

\begin{figure}[!ht]
  \begin{minipage}[t]{0.45\textwidth}
    \centering
    \includegraphics[width=1\textwidth]{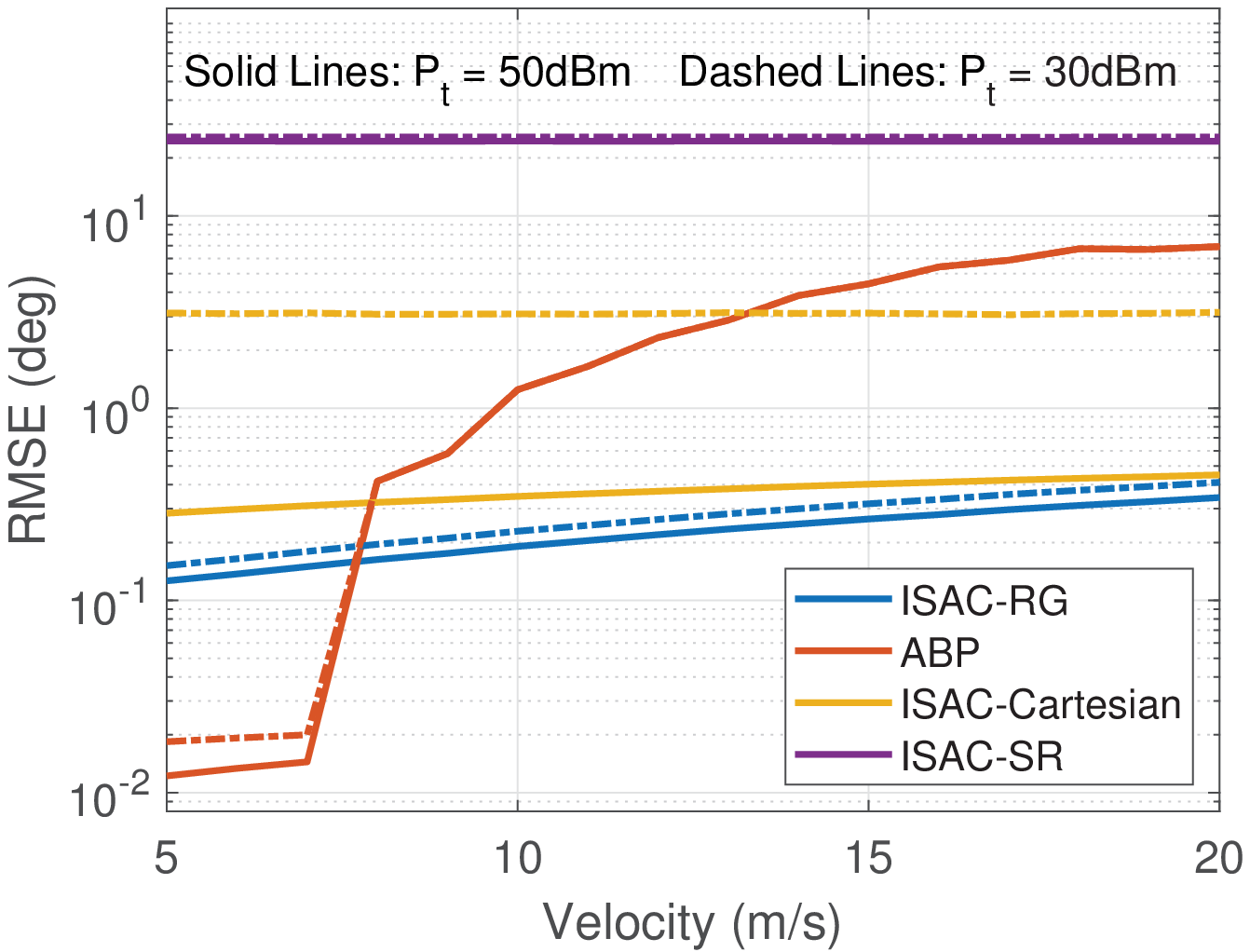}
    \caption{RMSE of the angle prediction versus the velocity of the vehicle in the country road model. }\label{RMSE_Velocity}
  \end{minipage}
  \quad
  \begin{minipage}[t]{0.45\textwidth}
    \centering
    \includegraphics[width=1\textwidth]{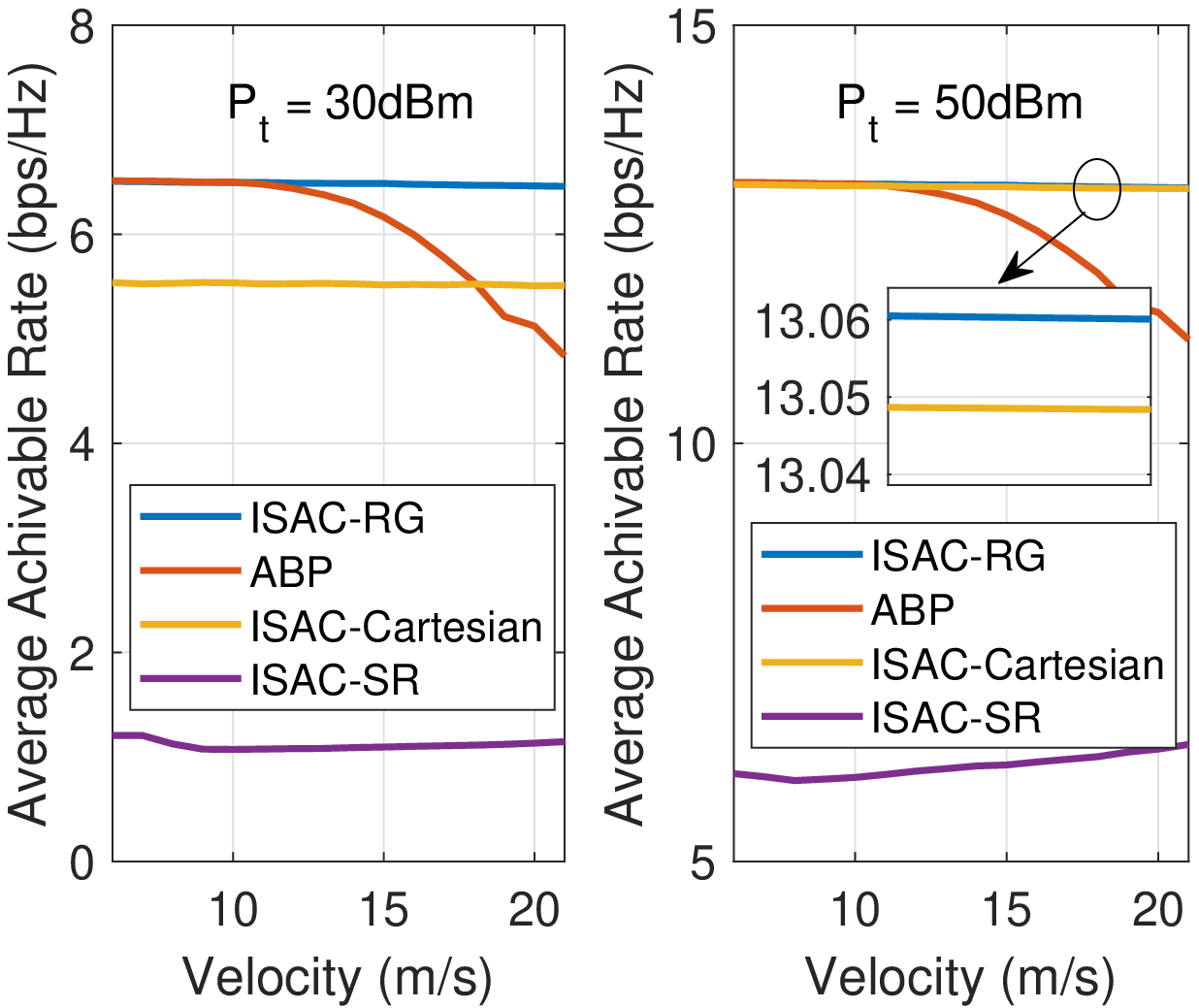}
    \caption{Average achievable rate versus the velocity of the vehicle in the country road model. }\label{ACR_Velocity}    
  \end{minipage}    
\end{figure}

To verify the superiority of the proposed method in high-mobility scenarios, we give the overall RMSE result of the angle prediction and the average achievable rate versus velocity in Fig. \ref{RMSE_Velocity} and Fig. \ref{ACR_Velocity}.
In general, the ISAC-RG and ISAC-Cartesian methods show less relationship between the tracking/communication performance and the velocity while the performance of the ABP algorithm degenerates quickly when the velocity is high enough.
As we mentioned above, the accuracy of the proposed EKF-based algorithm mainly relies on estimation accuracy in all past observations and the system noise, which is almost irrelative to how fast the vehicle moves.
For the ABP algorithm, the uncertainty imposed by the prediction is avoided by using the feedback information, which leads to better performance in the low-velocity regime. However, when the angular variation is too large to fall into the searching range, the ABP method quickly breaks down.
Meanwhile, since the ISAC-SR is not able to track the vehicle properly, the communication performance mainly relies on the power budget and how many times the beam aligns in accident while the tracking performance is almost meaningless. 

\subsection{Performance for IMM tracking and Behavior Reasoning}
\begin{figure}[!htbp]
  \begin{minipage}[t]{0.45\textwidth}
    \centering
    \includegraphics[width=1\textwidth]{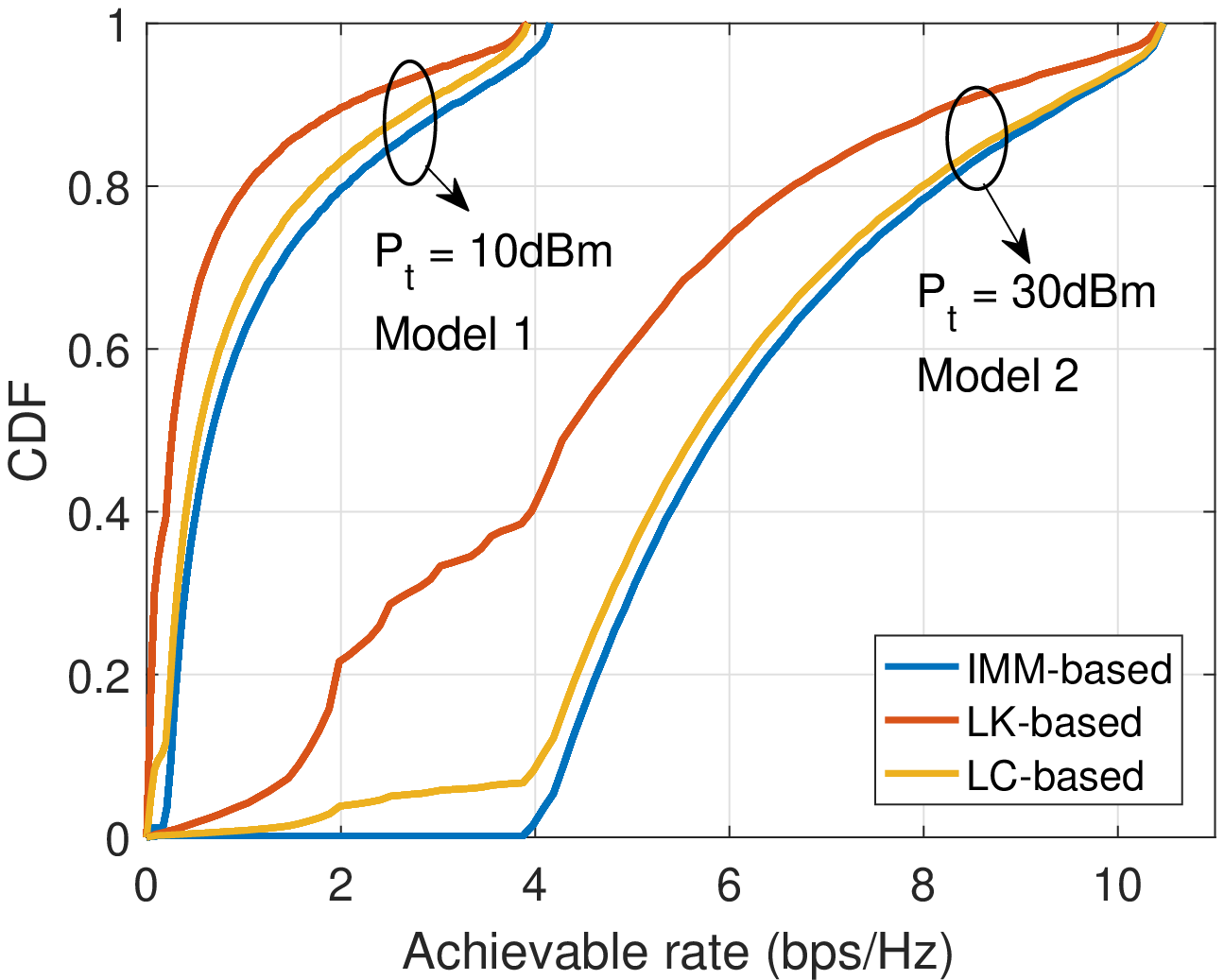}
    \caption{CDF of achievable rates for the IMM-based algorithm and single-model-based algorithms in the manuevere scenarios.}\label{IMM_ACR_CDF}    
  \end{minipage}
  \quad
  \begin{minipage}[t]{0.45\textwidth}
    \centering
    \includegraphics [width=1\textwidth]{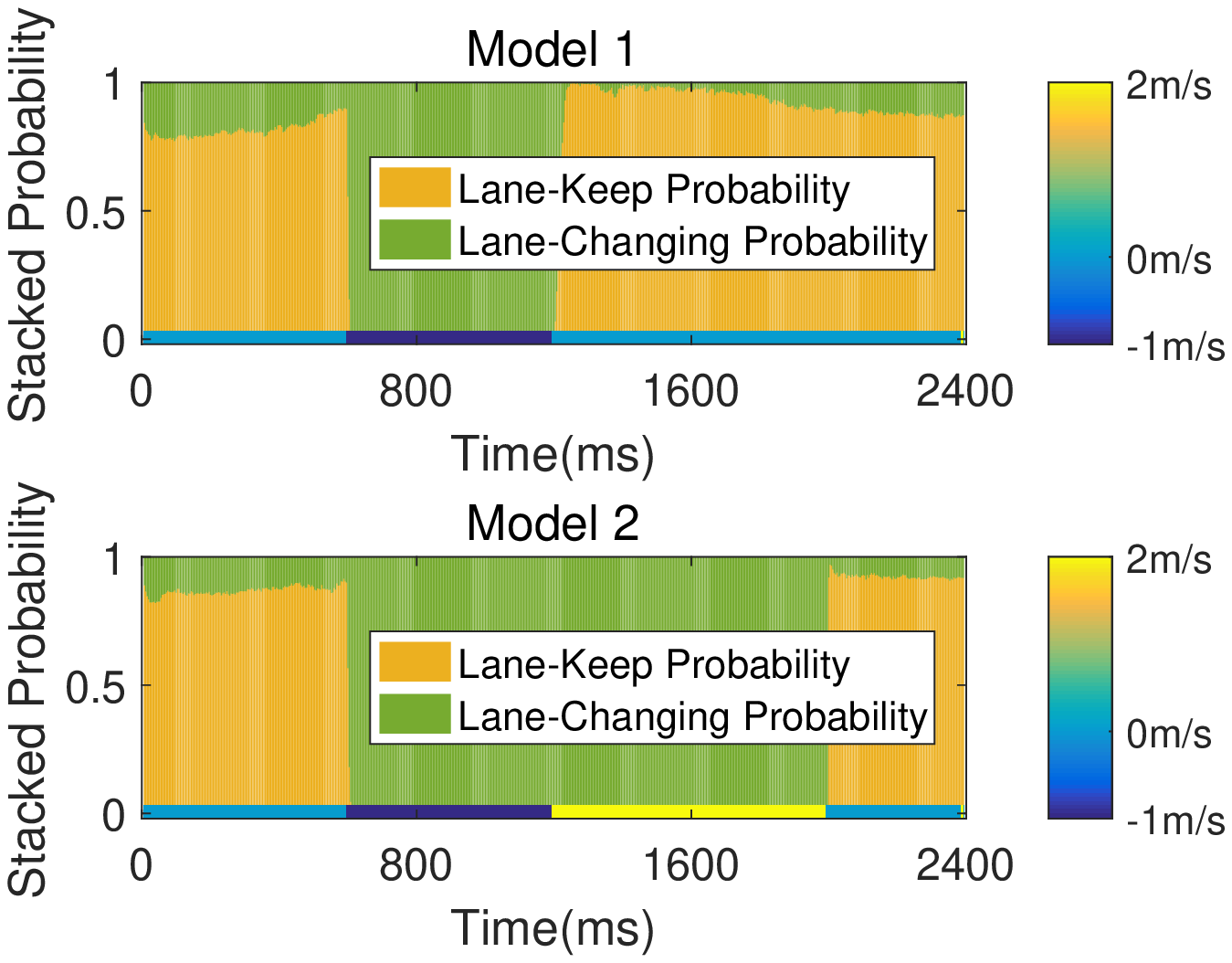}
   \caption{Reasoning Performance of IMM-based algorithm.}\label{IMM_Reasoning}
  \end{minipage}
\end{figure}

In this part, we study the performance of the proposed IMM-based tracking scheme.
We consider two scenarios where the vehicle tries to maneuver at a constant lateral velocity on the country road.
In the first scenario, the vehicle firstly moves stably at a longitudinal velocity of 5m/s from 0ms to  1200ms. Then the vehicle changes its lane from 1200ms to 2400ms at a lateral velocity of 1m/s while the longitudinal velocity is still 5m/s. After that, the vehicle continues to move stably.
In the second scenario, the vehicle has the same state as the first scenario before 2400ms but quickly changes its lane to the other direction in 2m/s from 2400ms to 3600ms.
The lateral velocity in the two scenarios is demonstrated as the bottom bar in Fig. \ref{IMM_Reasoning}.

In Fig. \ref{IMM_ACR_CDF}, we investigate the comparison of communication performance between the IMM-based algorithm and the single-model-based algorithms by showing the CDF versus achievable rates.
It is obvious that the IMM-based shows superiority over the single-model-based algorithms in the given scenarios.
More specifically, the achievable rate of the LK-based algorithm always has a notable gap between the other schemes which verifies that the EKF with the LK model is difficult to generalize to other models.
The performance of LC model also lags behind the IMM-based algorithm, especially in scenario 1 in which the vehicle only slightly changes lane.
It is because the LC model introduces an additional state variable to track the lateral movement and leads to more system noise and larger error of angle prediction, even though the vehicle is moving stably.

Then we show the stacked probability of the two models in Fig. \ref{IMM_Reasoning} to evaluate the reasoning ability of the IMM-based algorithm.
As shown in both the subfigures, when the lateral velocity is not zero, the probability of the LC model is close to 1, which indicates that we can always identify the lane-changing operation.
In the other situations, the probability of the LK model is not always close to 1 which slightly impairs the tracking performance. However, the reasoning performance is still guaranteed since the LK probability is larger than 0.5, which is usually regarded as a threshold of identification.
\footnote{Note: It is worth mentioning that we assume the lateral velocity is a constant when the vehicle changes its lane and mutated when the changing is stopped. This is an extreme case that barely occurs on the real-world road but can prove that the proposed algorithm is able to work even in extreme cases.}
\subsection{Performance for Dynamic Beamforming}
In this subsection, we compare the misalignment probability and achievable rate between the general constant beamwidth (CB) algorithm which tries to form the narrowest beam and our proposed dynamic beamwidth (DB) scheme which guarantees a given misalignment probability.
To be mentioned, we assume the system noise as $\sigma_{s} = 0.16m$ and $\sigma_{n} = 0.0032m$ to reveal the performance under a worse situation where the error of the roadway geometry map is larger or the behavior of the vehicle is more unpredictable.

\begin{figure}[!htbp]
  \begin{minipage}[t]{0.45\textwidth}
    \centering
    \includegraphics[width=1\textwidth]{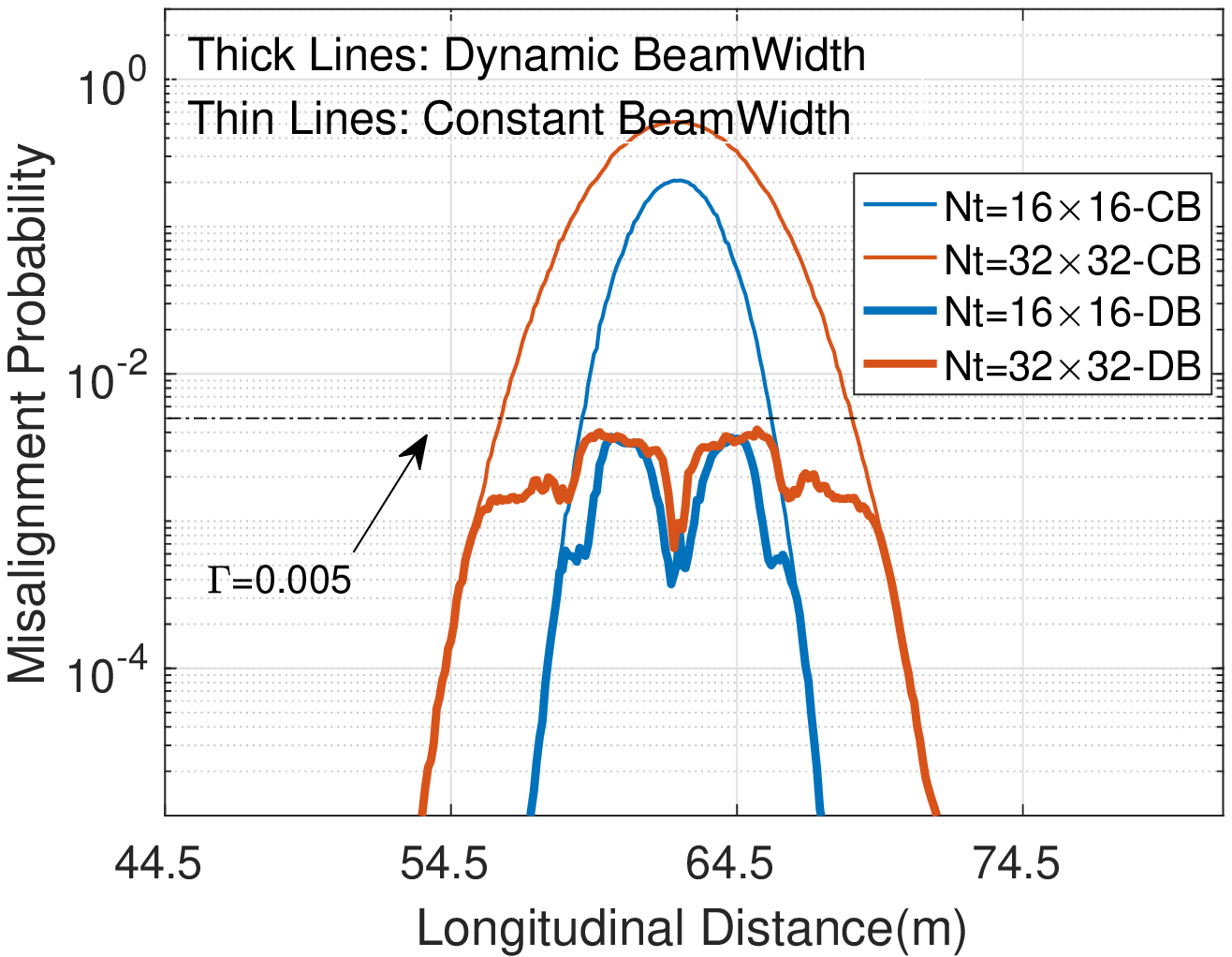}
    \caption{Misalignment probability of constant beamwidth scheme and proposed dynamic beamwidth shceme when the vehicle is near to the RSU in country road model.}\label{Outage_DB}
  \end{minipage}
  \quad
  \begin{minipage}[t]{0.45\textwidth}
    \centering
    \includegraphics[width=1\textwidth]{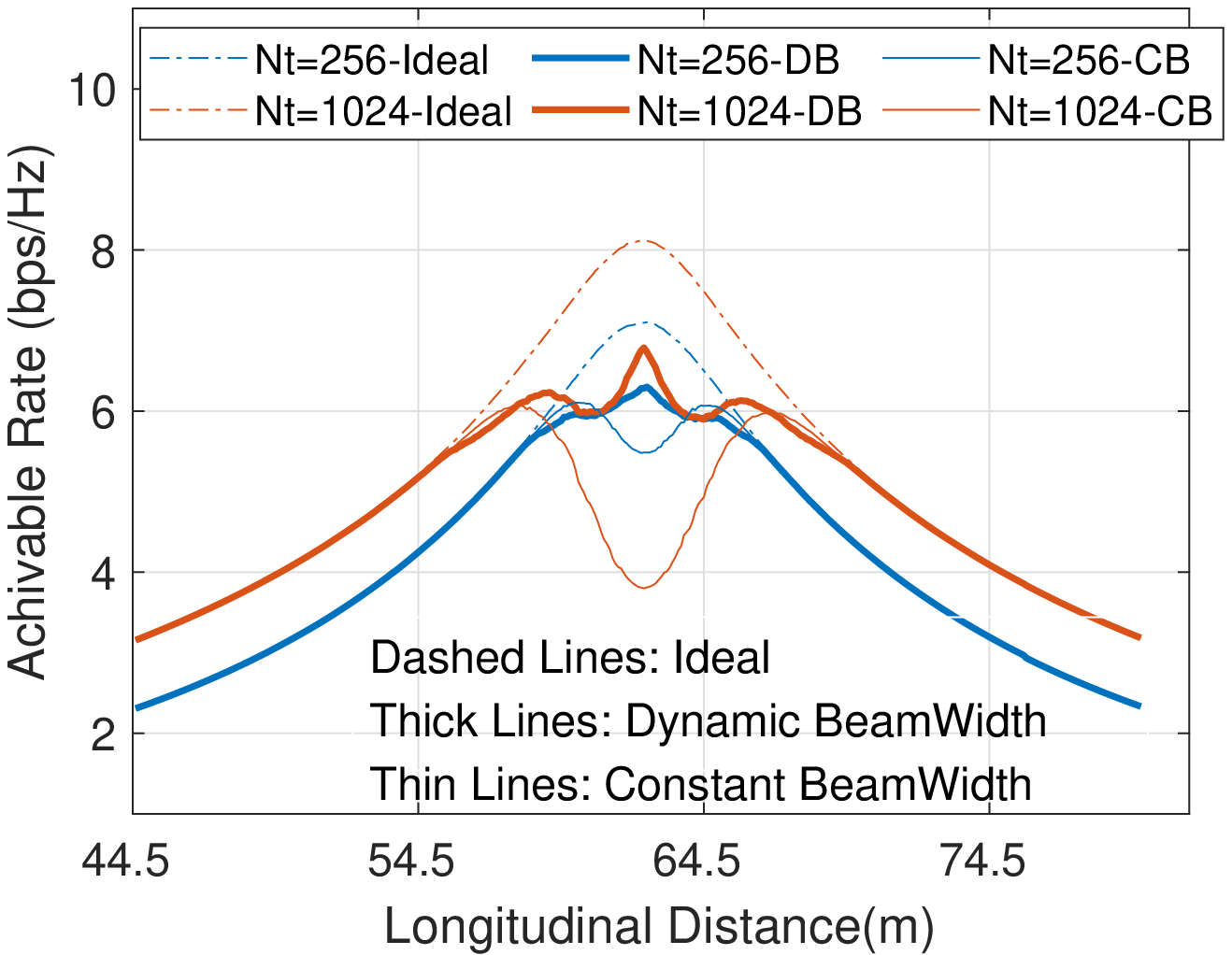}
    \caption{Achievable rate of constant beamwidth scheme and proposed dynamic beamwidth shceme when the vehicle is near to the RSU in country road model.}\label{ACR_DB}
  \end{minipage}
\end{figure}

In Fig. \ref{Outage_DB}, we first evaluate the misalignment probability of the two schemes. 
It can be observed that the misalignment probability of the CB algorithm increases with respect to the size of the antenna array while that of the DB algorithm is always lower than the given threshold.
Then we focus on the tendency of probability to change along the positional axis.
The change of the CB algorithm is obviously intuitive, where the misalignment probability increases with the vehicle nearing the RSU.
It is worth pointing out that there is a distinct gap between the threshold and the actual misalignment probability of the DB scheme at some certain points. 
The first reason of this phenomenon is that when the size of the antenna is small, the discreteness of adjustment makes it impossible to achieve the best beam-covered area and accordingly leads to lower misalignment probability and potential performance loss of achievable rate.
The second reason is that when the cross-correlation between the two spatial frequencies is large, the rotation angle of the eclipse of the PDF is approximated to be $\pi/4$ while that of the beam-covered area is always $0$, which results in the largest residual error of the approximation.

We then compare achievable rates of the proposed scheme, the CB scheme and the ideal scenario where the system always has a full array gain in Fig.\ref{ACR_DB}.
Both the algorithms show the same performance with the ideal scenario when the vehicle is far from the RSU where the misalignment probability is low enough which is consistent with the analysis above.
As the misalignment of DB algorithm is lower, the communication performance shows superiority over the DB algorithm and inferiority as compared with the ideal scenario.
In addition, it is interesting to note that crossover occurs in the area where the performance curves begin to separate, which is because the discreteness of the array size and the residual error leads to excessive adjustment and performance loss.
Fortunately, the degeneration is moderate and may not occurs in other circumstance.

\begin{table}[!htbp]
  \centering
  \caption{computational complexity for different algorithms per data frame.}
\begin{tabular}{|c|c|c|c|c|}
  \hline
  \quad&ISAC-RG&ISAC-Cartesian&ABP&ISAC-SR\\
  \hline
  Analytical   &$\mathcal{O}(K\zeta)$&$\mathcal{O}(K)$&$\mathcal{O}(K)$&$\mathcal{O}(K)$\\
  \hline
  Numerical ($\mu$s)&53.29&8.25&4.73&21.73\\
  \hline
\end{tabular}
\end{table}

To verify the practical feasibility of introducing the IMM-EKF-based algorithm with dynamic beamwidth adjustment under the CCS, we give a brief discussion on the complexity of our proposed method.
The number of the VU $K$ is assumed to be 1 and the number of the potential kinematic $\zeta$ is assumed to be 2 in this simulation.
The simulation is performed on an Intel Core i5-11500T CPU 32GB RAM computer with 1.5GHz.
As shown in Table 1, the overheads of the benchmarks are all $\mathcal{O}(K)$ while that of the proposed method is $\mathcal{O}(K\zeta)$. The additional overhead comes from the IMM algorithm which simultaneously tracks multiple kinematic models.
However, since there is no iterative procedure in the proposed method (neither in the benchmarks), the average per-frame execution time on the PC is 53.29 $\mu$s, which can be greatly reduced on a 5G cellular base station.
Moreover, the computing tasks for the elementary filters are irrelative to each other, which means while complexity is $\mathcal{O}(K\zeta)$, the execution time can be reduced to $\mathcal{O}(K)$ by employing parallel computing techniques.

\section{Conclusion}
In this article, we have proposed a novel predictive beamforming scheme for tracking and communicating with a vehicle on arbitrarily shaped roads by employing an RSU equipped with the ISAC capability.
By applying the CCS, the proposed approach enabled the implementation of high accuracy EKF beam prediction to detect and track the vehicle given any roadway geometries.
Furthermore, we have proposed an IMM-based filtering scheme to track and identify vehicle maneuvering.
Considering the beam misalignment incurred by system noise and measurement noise, a dynamic beamwidth adjustment algorithm has been further proposed by analyzing the properties of the spatial frequency uncertainty.
The algorithm is aimed at maximizing the array gain while ensuring that the misalignment probability is lower than a given threshold.
Finally, to validate the effectiveness of our proposed algorithms, 
numerical results have been provided to show that the IMM-EKF-based predictive beamforming scheme outperforms conventional benchmark techniques.
The ability of tracking and identifying the vehicle maneuvering has also been verified.
Additionally, the trade-off between the misalignment probability and the array gain has been demonstrated by adopting the dynamic beamwidth scheme.




\ifCLASSOPTIONcaptionsoff
  \newpage
\fi



\bibliographystyle{IEEEtran}
\bibliography{IEEEabrv,CEP_REF,EKF_RG}
\end{document}